\documentclass[graybox, envcountchap]{svmult}

\usepackage{mathptmx}        
\usepackage{helvet}          
\usepackage{courier}         

\usepackage{makeidx}         
\usepackage{graphicx}        
\usepackage{multicol}        
\usepackage[bottom]{footmisc}

\makeindex             

\usepackage{amsmath}
\usepackage{amssymb}
\usepackage{lscape}
\usepackage{multirow}

\def\gapp{\ifmmode\stackrel{>}{_{\sim}}\else$\stackrel{<}{_{\sim}}$\fi}
\def\gsim{\lower.5ex\hbox{\gtsima}}
\def\gtsima{$\; \buildrel > \over \sim \;$}
\def\lapp{\ifmmode\stackrel{<}{_{\sim}}\else$\stackrel{<}{_{\sim}}$\fi}
\def\lsim{\lower.5ex\hbox{\ltsima}}
\def\ltsima{$\; \buildrel < \over \sim \;$}

\newcommand\apgt{\ {\raise-.5ex\hbox{$\buildrel>\over\sim$}}\ }
\newcommand\aplt{\ {\raise-.5ex\hbox{$\buildrel<\over\sim$}}\ }
\newcommand{\arcsec}     {\mbox{\ensuremath{{}^{\prime\prime}}}}%

\begin{document}
\pagestyle{empty}
\frontmatter


\mainmatter

\setcounter{chapter}{4}
\titlerunning{BSSs in globular clusters: observations}
\authorrunning{Ferraro et al.}

\title{Blue Straggler Stars in Globular Clusters: a powerful tool to
  probe the internal dynamical evolution of stellar systems}

\author{Francesco R. Ferraro, Barbara Lanzoni, Emanuele Dalessandro, Alessio Mucciarelli, and Loredana Lovisi}
\institute{Francesco R. Ferraro et al. \at Dipartimento di Fisica e Astronomia,
  Universit\`a degli Studi di Bologna, Via Ranzani 1,
  I--40127 Bologna, Italy, \email{francesco.ferraro3@unibo.it}
 }

\maketitle
\label{Chapter:Ferraro}

\abstract*{
This chapter presents an overview of the main observational results
obtained to date about Blue Straggler Stars (BSSs) in Galactic
Globular Clusters (GCs).  The BSS specific frequency, radial
distribution, chemical composition and rotational properties are
presented and discussed in the framework of using this stellar
population as probe of GC internal dynamics. In particular, the shape
of the BSS radial distribution has been found to be a powerful tracer
of the dynamical age of stellar systems, thus allowing the definition
of the first empirical ``dynamical clock''.
}

\section{Introduction}
\label{fersec:intro}
\begin{figure}
\begin{center} 
\includegraphics[width=119mm]{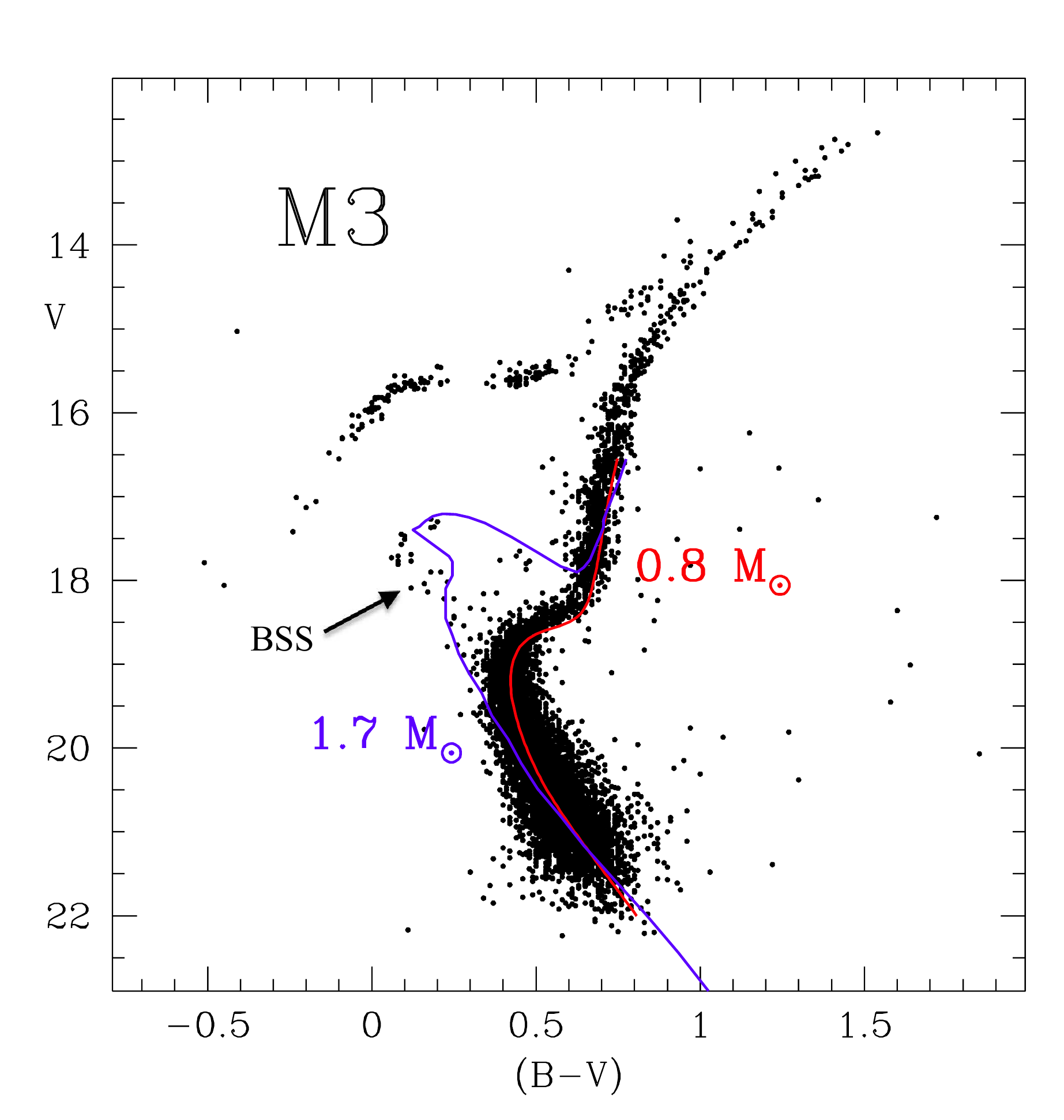}
\caption{Optical CMD of the globular cluster M3, with the location of
  BSSs indicated by the arrow. The theoretical track corresponding to
  $0.8$ M$_\odot$  reproduces well the main evolutionary sequences of the
  cluster, while BSSs populate a region of the CMD where core
  hydrogen-burning stars of $\sim 1.7$ M$_\odot$ are expected. From
  \cite{buon94}.}
\label{m3}
\end{center}
\end{figure}

Blue straggler stars (BSSs) in globular clusters\index{globular cluster} (GCs) are commonly
defined as those stars located along an extrapolation of the main
sequence (MS), in a region brighter and bluer (hotter) than the
turnoff (TO) point, in the optical colour-magnitude diagram\index{colour-magnitude diagram} (CMD; see
Fig.~ \ref{m3}).  They were first discovered by Sandage \cite{sand53} in the
external region of the Galactic GC M3\index{M3}. Their origin has been a mystery
for many years and the puzzle of their formation is not completely
solved yet.

The BSS location in the CMD suggests that they are more massive than
the current cluster population (Fig.~\ref{m3}; this is also
  confirmed by a few mass measurements, e.g., \cite{shara97,gilliland98}).
However, GCs are completely devoid of gas and any recent star
formation event can be realistically ruled out.  Hence the BSS origin
should be searched in some mechanisms able to increase the initial
mass of single stars in a sort of {\it rejuvenation process}. The BSS
formation mechanisms are not completely understood yet, but the main
leading scenarios, at present, are mass transfer\index{mass transfer} (MT) processes
between binary companions \cite{mcrea64,zinnsearle76}, possibly up to
the complete coalescence\index{coalescence} of the binary system, or the merger\index{merger} of stars
induced by collisions\index{collision} (COLL; \cite{hillsday76}).  Being more
massive than the average cluster stars, BSSs suffer from the effect of
dynamical friction\index{dynamical friction}, that makes them sink towards the cluster centre
\cite{mapelli04,fe12}.  In turn, the frequent stellar interactions
occurring in the ultra-dense cores of Galactic GCs can promote both
the formation and the hardening of binary systems, thus contributing
to generate MT-BSSs. All these considerations clearly show that BSSs
represent a crucial link between standard stellar evolution and GC
internal dynamics (see \cite{bai95,fe12}, and references therein).

\section{BSS Specific Frequency and Primordial Binary Fraction} 
 \label{fersec:freq}
 
The observational and interpretative scenario of BSSs has
significantly changed in the last years. In fact, for a very long time
(almost 40 years) since their discovery (see Chap. 2), 
BSSs have been detected only
in the outer regions of GCs, or in relatively loose clusters. This
generated the idea that low-density environments were their {\it
  natural habitats}.  However, starting from the early '90, high
spatial resolution facilities allowed to properly image and discover
BSSs also in the highly-crowded central regions of dense GCs (see the
case of NGC\,6397\index{NGC 6397}; \cite{auriere90}), thus demonstrating that the
previous conviction was just due to an observational bias.  In
particular, the advent of the \emph{Hubble Space Telescope}\index{Hubble Space Telescope} (HST) represented
a real turning point in BSS studies, thanks to its unprecedented
spatial resolution and imaging/spectroscopic capabilities in the
ultraviolet\index{ultraviolet} (UV). Indeed, the pioneering observations of the central
regions of 47 Tucanae\index{47 Tucanae} \cite{paresce91, guha94} and M15\index{M15}
\cite{feparesce93}, by means of the HST opened a new perspective in
the study of BSSs, definitely demonstrating that they also
(preferentially) populate high-density environments.
   
Based on these observations, the first catalogs of BSSs have been
published (e.g., \cite{fp92,saraj93}) and the first comparisons
among different clusters have been attempted \cite{fe95}. In
order to perform meaningful comparisons, various definitions of BSS
specific frequencies have been proposed over  time.  Ferraro and colleagues \cite{fe93}
introduced the ``double normalised ratio'', defined as:
\begin{equation}
 R_{\rm BSS} = {{(N_{\rm BSS}/N_{\rm BSS}^{\rm tot})} \over {(L^{\rm
       sampled}/L_{tot}^{\rm sampled})}},
\label{eq:rbss}
\end{equation}
where $N_{\rm BSS}$ is the number of BSSs counted in a given cluster
region, $N_{\rm BSS}^{\rm tot}$ is the total number of BSSs observed,
and $L^{\rm sampled}/L_{tot}^{\rm sampled}$ is the fraction of light
sampled in the same region, with respect to the total measured
luminosity. The same ratio can be defined for any post-MS population.
Theoretical arguments \cite{renzfusi88} demonstrate that the double
normalised ratio is equal to unity for any population (such as red
giant\index{red giant} branch and horizontal branch stars\index{horizontal branch star}, RGB and HB, respectively)
whose radial distribution follows that of the cluster
luminosity. Other definitions of the BSS specific frequency adopted in
the literature are: $S4_{\rm BSS}=N_{\rm BSS}/L_s$, where $N_{\rm BSS}$
is the number of BSSs and $L_s$ is the sampled luminosity in
units of $10^4$ L$_\odot$ \cite{fe95}; $F_{\rm BSS}=N_{\rm BSS}/N_{\rm
  bright}$, where $N_{\rm bright}$ is the number of all the stars
brighter than two magnitudes below the HB level \cite{bolte93};
$F^{\rm BSS}_{\rm pop}=N_{\rm BSS}/N_{\rm pop}$ \cite{fe03sixGCs},
where $N_{\rm pop}$ is the number of stars belonging to a cluster
``normal'' population adopted as reference (generally the HB population,
or a segment of the RGB or MS).  Besides the different definitions,
all these normalisations account for the different cluster richness
(i.e, total luminosity or mass). However, as discussed in \cite{fe95},
particular caution is needed when looking for correlations among BSS
specific frequencies and cluster structural parameters, since the
concentration parameter and central density are intrinsically related
to the cluster luminosity (see, e.g., \cite{djorgmey93}); hence
spurious correlations can emerge simply because of the BSS
``normalisation''.

The largest compilations of BSSs to date have been collected for
nearly 60 Galactic GCs surveyed with the HST/WFPC2
\cite{piotto04, leigh07, moretti08}, and for 35 clusters
\cite{leigh11} observed within the HST/ACS \emph{Survey of Galactic
Globular Clusters} \cite{saraj07}.  These compilations, together with
deep investigations in open clusters, dwarf spheroidals, and the
Galactic field (see Chap. 3, 4, and 6 in this book), have significantly contributed to
form the nowadays largely accepted idea that BSSs are a stellar
population common to any stellar system.

\begin{figure}
\begin{center} 
\includegraphics[width=119mm]{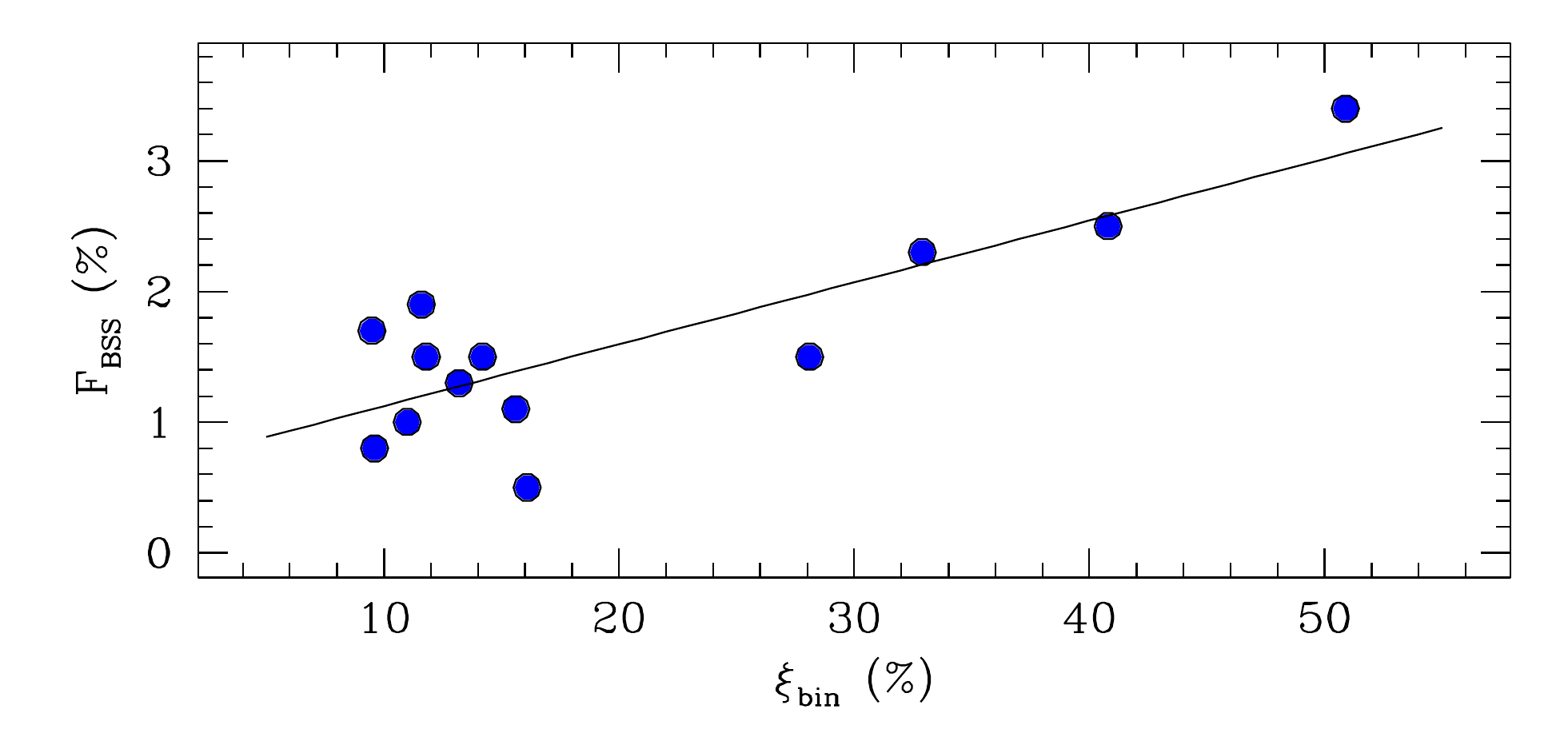}
\caption{BSS specific frequency as a function of the core binary
  fraction measured in a sample of low-density GCs. The best-fit
  linear correlation is also plotted (solid line). From
  \cite{sollima08}.}
\label{bssbin} 
\end{center}
\end{figure}

With the aim of understanding how BSSs form and if their formation
mechanisms depend on some cluster physical properties, the most recent
catalogs have been used to search for correlations between the BSS
specific frequency (or number), and several parameters tracing the
cluster structure (as luminosity, mass, central density, etc.), as
well as for correlations with the collision rate and binary fraction
(for recent results, see Chap. 9).  Though not conclusive, this approach has provided a number
of interesting results. For instance, no correlation has been found
with the collisional parameter
\cite{piotto04,davies04,leigh07}, while a strong
correlation has been revealed between the number of BSSs in cluster
cores and the core mass\index{cluster core mass} \cite{knigge09,leigh13}.  These facts have
been interpreted as the evidence of a binary (instead of a
collisional) origin of BSSs, even in the densest environments, like
the centre of post-core collapsed (PCC)\index{post-core collapse} clusters \cite{knigge09}.
However, the fraction of binary systems in a sample of 59 GCs has been
recently estimated from the distribution of stars along the
``secondary'' MS \cite{milone12}, thus allowing to explore directly
any possible correlation between the fraction/number of BSSs and that
of binaries.  By using a sub-sample of 24 GCs, Milone et al. \cite{milone12} found
a nice correlation between the BSS specific frequency and the binary
fraction\index{binary fraction} in cluster cores.  This has been confirmed also by
Leigh et al. \cite{leigh13}, who, however, obtain a much stronger correlation
between the number of core BSSs and the cluster core mass.
Interestingly, in Milone et al. \cite{milone12} plot, PCC clusters lie well
outside the relation.  This likely reflects the role that internal
dynamics plays on the binary and BSS content of GCs.  In fact, binary
systems are subject to frequent dynamical interactions with other
binaries, single stars and even multiple systems. These interactions
can either bring to stellar collisions, or significantly alter the
physical properties of binaries, even promoting mass transfer
activity.  Hence, binaries and interactions play a crucial role in
both the MT and the COLL scenarios and it is probably impossible to
separate the two effects just on the basis of the observed binary
fraction.  An exception could be represented by low density
environments, where the efficiency of dynamical interactions is
expected to be negligible. Very interestingly, indeed, a clear
correlation between the binary and the BSS frequency has been found in
a sample of 13 low density GCs ($\log \nu_0<3$ in units of
  L$_\odot/$pc$^3$; see Fig.~\ref{bssbin} and \cite{sollima08}).  This is
the cleanest evidence that the unperturbed evolution of primordial
binaries is the dominant BSS formation process in low-density
environments (also consistently with the results obtained in
  open clusters; e.g., \cite{mathieu09}).

\section{The Ultraviolet Route to Study BSS}
\label{fersec:UV}
 
\begin{figure}
\begin{center} 
\includegraphics[width=119mm]{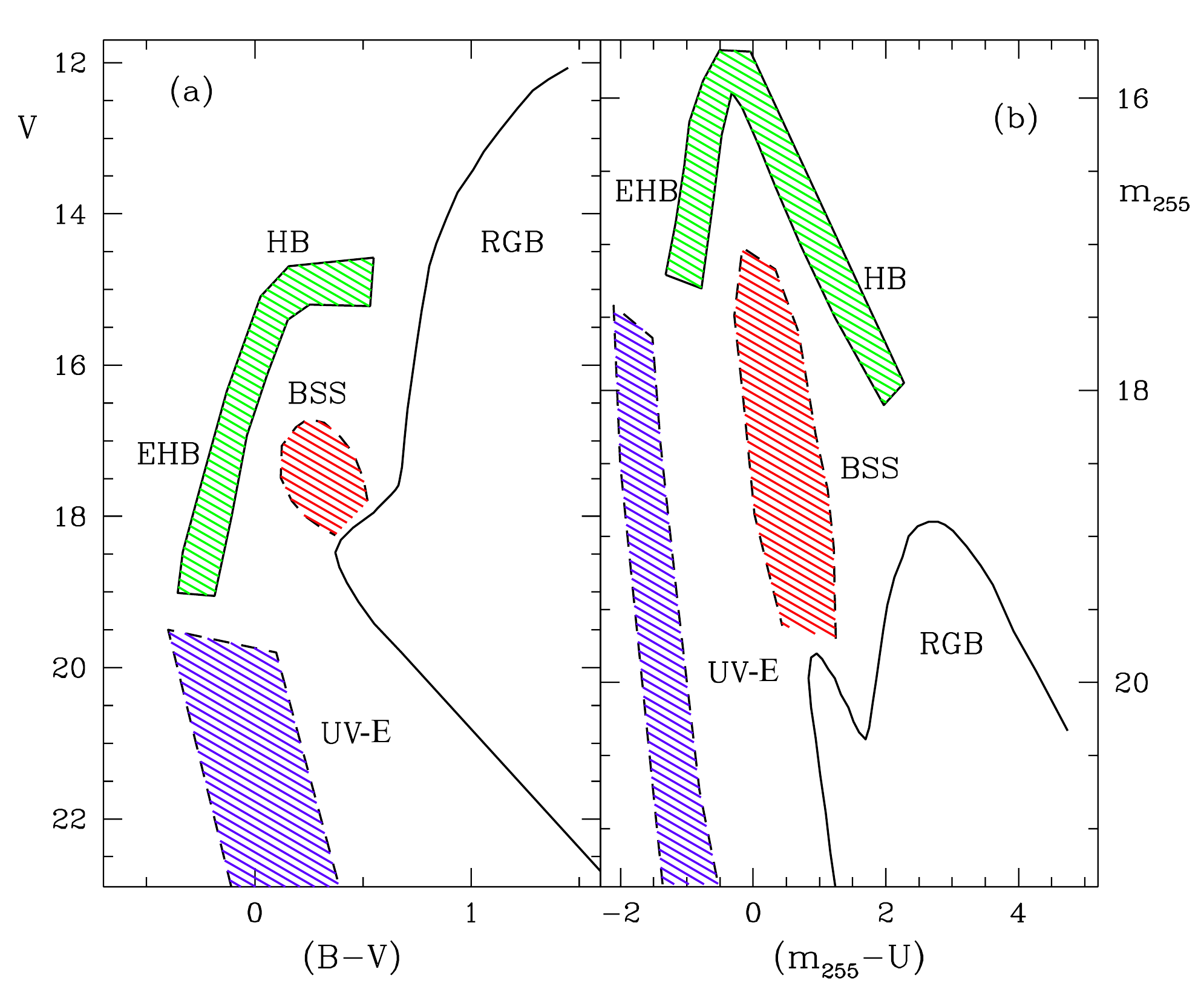}
\caption{Sketch of the stellar evolutionary sequences in
  the optical (left panel) and in the UV (right panel) CMDs. The loci
  of RGB, BSS, HB, extreme-HB stars (EHB) and stars with UV excess are
  marked.}
\label{uvcmd}  
\end{center}
\end{figure}

The systematic study of BSSs in the visible-light bands, especially in the
central regions of high density clusters, is intrinsically difficult
and remains problematic even when using HST.  This is because the
optical emission of old stellar populations is dominated by cool,
bright giants.  Hence, the observation of complete samples of (faint)
hot stars (as BSSs, other by-products of binary system evolution,
extreme blue horizontal branch stars, etc.) is quite problematic in
this plane. {\it It is like trying to make a complete census of
  fire-flies, while having a clump of large light-bulbs just in front
  of us. In order to secure a proper counting of fire-flies one needs
  to switch off the lights, first!}.  Moreover, in the visible-light plane,
BSSs can be easily mimicked by photometric\index{photometry} blends of subgiant branch
(SGB) and RGB stars.  Instead, at ultraviolet (UV)\index{ultraviolet} wavelengths RGB stars are very
faint, while BSSs are among the brightest objects.  In particular, in
the UV plane BSSs are much more easily recognisable, since they define
a narrow, nearly vertical sequence spanning a $\sim 3$ mag interval
(see Fig.~\ref{uvcmd}).  In the mean time, BSS-like blends are much
less severe at these wavelengths because of the relative faintness of
SGB and RGB stars.  Indeed, the ($m_{255}, \,m_{255}-m_{336}$) plane
has been found to be ideal for selecting BSSs even in the cores of the
densest GCs.

\begin{figure}
\begin{center} 
\includegraphics[width=119mm]{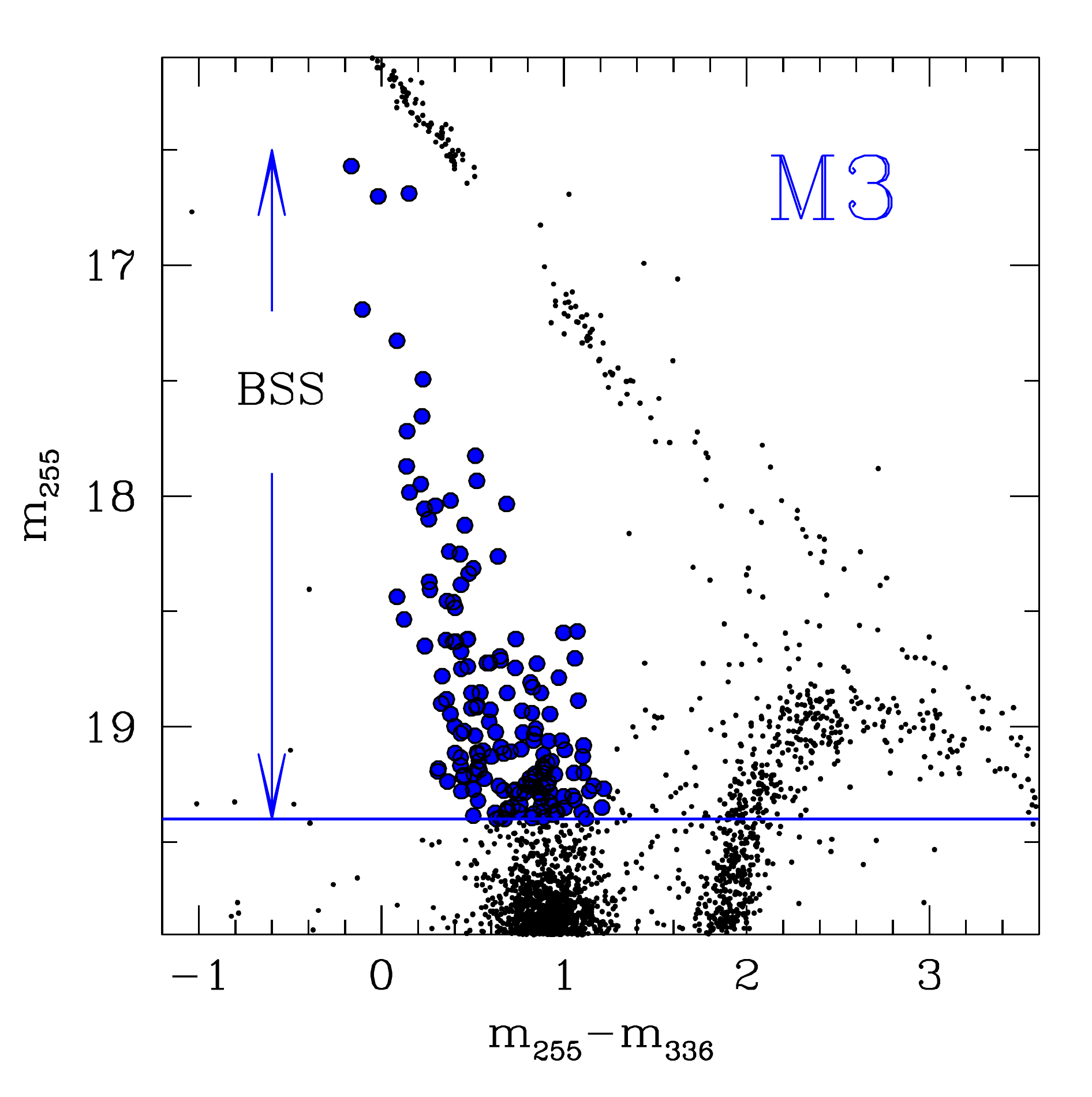}
\caption{BSS in the UV: the case of M3\index{M3}.  The horizontal line at
  $m_{255}=19.4$ is the assumed BSS limiting magnitude, corresponding
  to $\sim 5 \sigma$ above the turnoff level. From \cite{fe97}.}
\label{uvm3}
\end{center}
\end{figure}

\begin{figure}
\begin{center} 
\includegraphics[width=119mm]{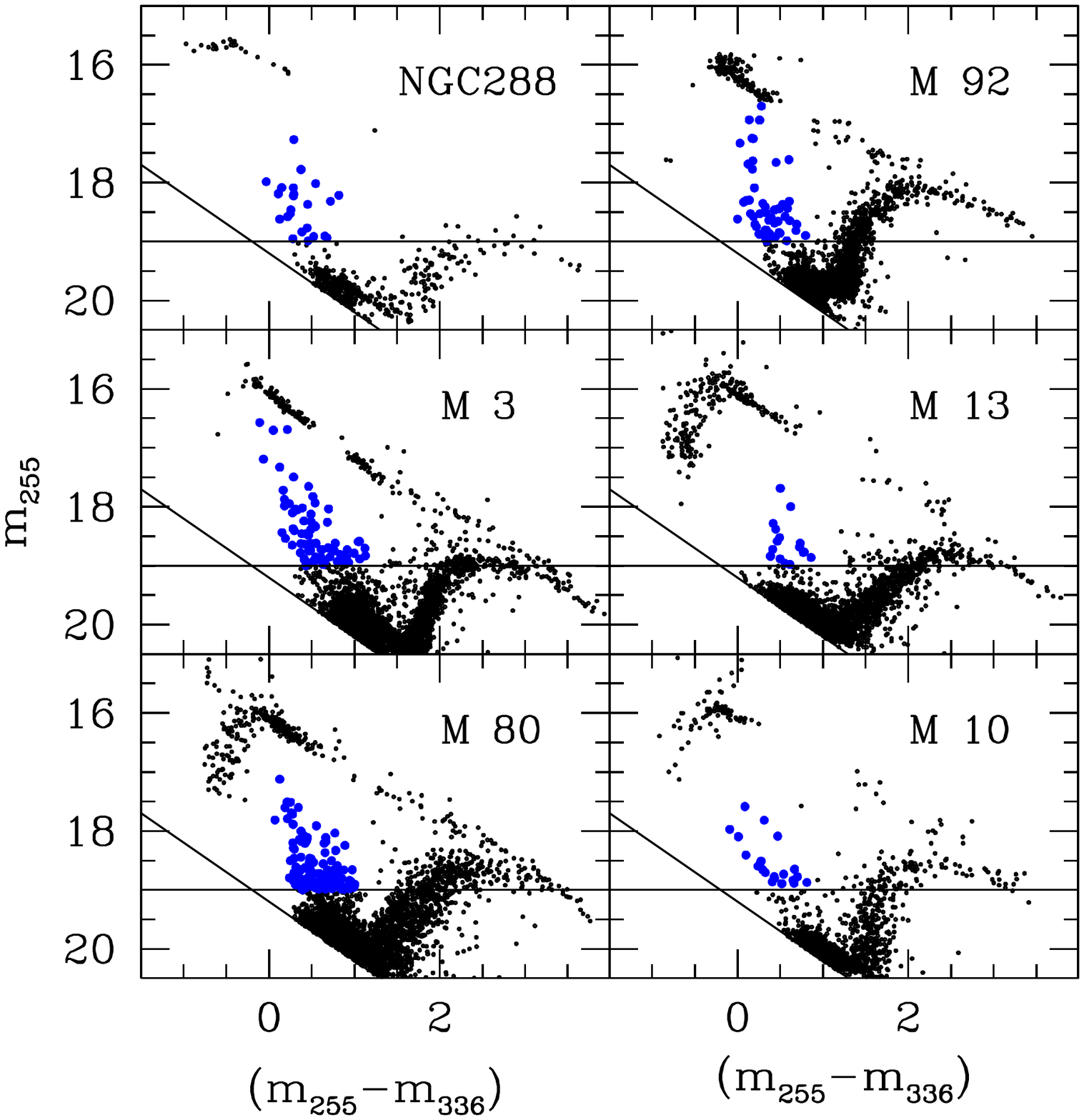}
\caption{UV CMDs of six different globular clusters (see
  labels), with the BSS populations highlighted as blue circles. The
  horizontal line marks the magnitude limit adopted for the BSS
  selection. From \cite{fe03sixGCs}.}
\label{uv6gcs}
\end{center}
\end{figure}

The UV exploration of the very central region of M3\index{M3} (see \cite{fe97}, hereafter
  F97; see also Fig.\ref{uvm3}] brought to the discovery of a
substantial population of BSSs, at odds with the depletion claimed by Bolte, Hesser \& Stetson
\cite{bolte93}. Since then, the central regions of a number of GCs
have been explored in the UV to search for BSSs: this systematic
approach allowed  putting the BSS study into a more quantitative basis
than ever before. Indeed a number of interesting results have been
obtained from cluster-to-cluster comparisons (see Fig.~\ref{uv6gcs}). Of course the UV approach strongly favours the
observation of hot objects, like HB stars (see Fig.~
\ref{uvcmd}). Hence the HB\index{horizontal branch star} becomes the ideal reference population for
the definition of the BSS specific frequency ($F_{\rm HB}^{\rm BSS}$)
at these wavelengths.
 
By using appropriate observations in the UV (Fig.~\ref{uv6gcs}), Ferraro et al.\cite{fe03sixGCs} presented a comparison of the BSS
populations in the central regions of 6 clusters (namely NGC\,288\index{NGC 288},
M92\index{M92}, M3\index{M3}, M13\index{M13}, M80\index{M80}, and M10\index{M10}) characterised by different structural
parameters.  The BSS specific frequency ($N_{\rm BSS}/N_{\rm HB}$) has
been found to largely vary from cluster to cluster, from 0.07 to 0.92,
and it does not seem to be correlated with central density, total
mass, velocity dispersion, or any other obvious cluster property
(see also \cite{piotto04}). On the other hand, this study pointed
out peculiar cases that statistical approaches (as those presented,
e.g., by \cite{piotto04}) did not bring into evidence. ``Twin''
clusters like M3 and M13 have been found to harbour quite different BSS
populations: the specific frequency in M13 is the lowest ever measured
in a GC (0.07), and it turns out to be 4 times lower than that
measured in M3 (0.28). \emph{What is the origin of this difference?}
We \cite{fe03sixGCs} suggested that it could be related to their binary
content; in particular the paucity of BSSs in M13 could be due either
to a quite poor population of primordial binaries, or to the fact that
most of them were destroyed during the cluster evolution.  Indeed the
fraction of binaries recently measured in the central regions of these
two clusters \cite{milone12} confirms a significant difference
($f_{\rm bin}=0.027$ for M3, and $f_{\rm bin}=0.005$ for M13), thus
supporting the hypothesis that this is the origin of the different BSS
content.  One of the most interesting result is that the largest BSS
specific frequency has been found in two GCs which are at the extremes
of the central density values in the analysed sample: NGC\,288 and
M80, with the lowest and the highest central density, respectively
($\log \rho_0=2.1$ and 5.8, in units of M$_\odot/$pc$^3$).  This
suggests that the two formation channels can have comparable
efficiency in producing BSSs in the respective most favourable
environment.
 
\begin{figure}
\begin{center} 
\includegraphics[width=119mm]{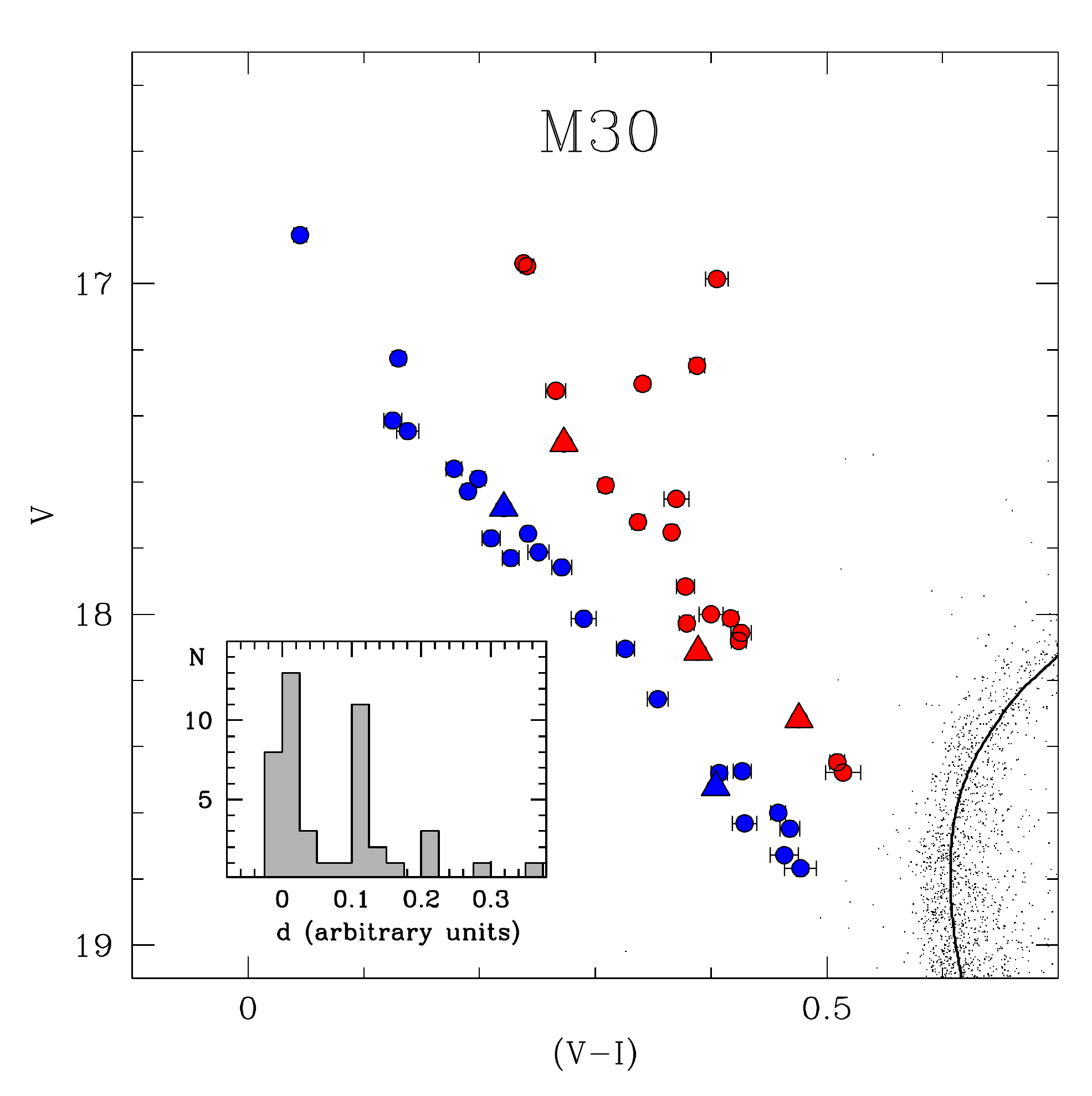}
\caption{Optical CMD of M30\index{M30} zoomed in the BSS
  region. The two distinct sequences of BSSs are highlighted as blue
  and red symbols. The inset shows the distribution of the geometrical
  distances of BSSs from the straight line that best fits the blue BSS
  sequence. Two well-defined peaks are clearly visible, confirming
  that the two sequences are nearly parallel to each other. From
  \cite{fe09m30}.}
\label{double} 
\end{center}
\end{figure}

\begin{figure}
\begin{center} 
\includegraphics[width=119mm]{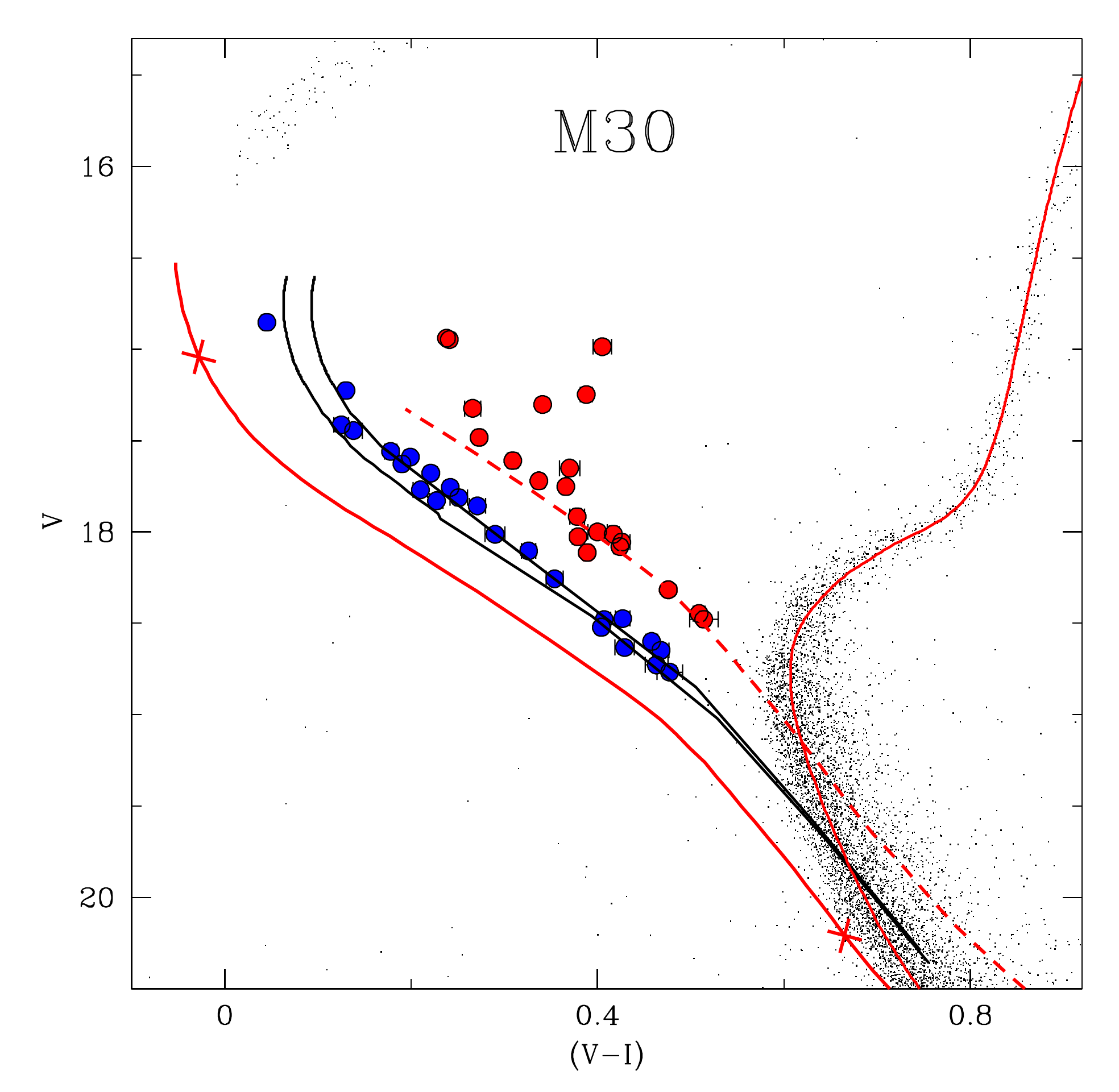} 
\caption{Magnified portion of the CMD of M30\index{M30}. The solid
  black lines correspond to the collisional isochrones of 1 and 2 Gyr,
  respectively, which accurately reproduce the blue BSS sequence. The
  solid red lines correspond to the single-star isochrones of 13 Gyr
  (well fitting the main cluster evolutionary sequences) and 0.5 Gyr
  (representing the reference cluster zero-age main sequence,
  ZAMS). The two crosses mark the respective positions of a 0.8
  M$_\odot$ star and a 1.6 M$_\odot$ star along the ZAMS. The dashed
  red line corresponds to the ZAMS shifted by 0.75 mag, marking the
  lower boundary of the locus occupied by mass-transfer\index{mass transfer} binary
  systems. This line well reproduces the red BSS sequence. From
  \cite{fe09m30}.}
\label{double_isocr}
\end{center}
\end{figure}

\begin{figure}
\begin{center} 
\includegraphics[width=119mm]{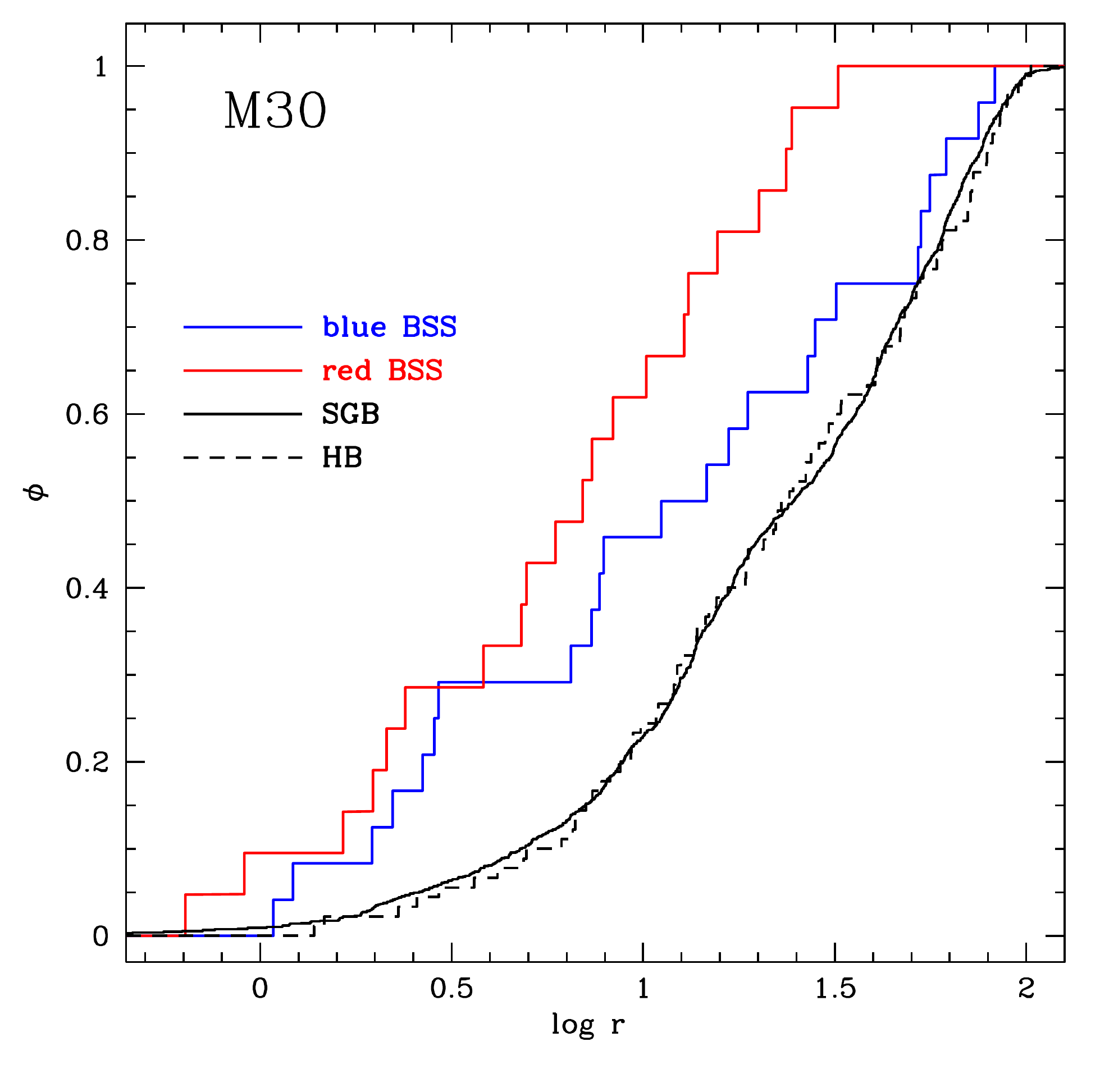}  
\caption{Cumulative radial distributions of red BSSs (red line) and
  blue BSSs (blue line), as function of the distance from the cluster
  centre. The distributions of subgiant\index{subgiant} branch stars (solid black
  line) and horizontal-branch stars\index{horizontal branch star} (dashed black line) are also
  plotted, for comparison. From \cite{fe09m30}.}
\label{double_cumu}
\end{center}
\end{figure}

\section{The Discovery of the Double BSS Sequence}
\label{fersec:double}

While the proposed BSS formation mechanisms could be separately at
work in clusters with different densities \cite{fe95,fe99}, a few
pieces of evidence are now emerging suggesting that they could also
act simultaneously within the same cluster. Indeed the discovery of a
double BSS sequence in M30\index{M30} (\cite{fe09m30}, hereafter F09) indicates
that this can be the case.
 
By using an exceptional set of 44 high-resolution images obtained with
the HST-WFPC2\index{Hubble Space Telescope}, F09 obtained a very high-precision CMD of the central
region of the Galactic GC M30.  The CMD revealed the existence of two
well-separated and almost parallel sequences of BSSs (Fig.~
\ref{double}). The two sequences are similarly populated, consisting
of 21 and 24 stars, respectively.  {\it This is the very first time
  that such a feature has been detected in any stellar system, and it
  could be the signature of the cluster core collapse imprinted onto
  the BSS population.}
  
The comparison with evolutionary models of BSS formed by direct
collisions\index{collision} of two MS stars \cite{sills09} shows that the blue-BSS
sequence is well fit by collisional isochrones\index{isochrone} with ages of $1-2$ Gyr
(black solid lines in Fig. \ref{double_isocr}).  Instead, the red-BSS
population is far too red to be properly reproduced by collisional
isochrones of any age, and its origin should therefore be
different. Binary evolution models \cite{tian06} have shown that
during the mass-transfer\index{mass transfer} phase (which can last several Gyr, i.e., a
significant fraction of the binary evolution time-scale), the binary
population defines a sort of ``low-luminosity boundary'' located $\sim
0.75$ mag above the zero-age MS in the BSS region. This is just where
the red-BSS sequence is observed (red dashed line in
Fig. \ref{double_isocr}). Hence, the BSS along the red-sequence could be
binary systems still experiencing an active phase of mass-exchange\index{mass exchange}.
 
Due to the normal stellar evolution, all BSSs will evolve toward the
RGB phase. In particular, the evolved blue-BSSs will populate the
region between the two observed sequences and fill the gap. Hence, the
fact that two well-separated chains of stars are observed supports the
hypothesis that both the blue- and the red-BSS populations have been
{\it generated by a recent and short-lived event, instead of a
  continuous formation process}. Quite interestingly, M30 is
classified as a PCC\index{post-core collapse} cluster in the original compilation of
Djorgovski \& King \cite{djorgking86}, and F09 confirmed this finding by carefully
re-determining the cluster density profile from deep HST images and
detecting a steep power-law cusp in the innermost 5\arcsec--6\arcsec
($\sim 0.2$ pc).  During the core collapse\index{core collapse} phase the central stellar
density rapidly increases, bringing to a concomitant enhancement of
gravitational interactions (in fact, the collisional parameter scales
as $\Gamma\propto \rho_0^{1.5}\,r_c^2$, where $r_c$ is the core
radius). In turn, these can trigger the formation of new BSSs, both
via direct stellar collisions and via mass transfer activity in
dynamically shrunk binary systems. All together these considerations
support a scenario where the two observed BSS populations are
generated by the same dynamical event (the core collapse): the
blue-BSSs arose from the enhanced stellar collision activity, while
the red-BSSs are the result of the evolution of binary systems which
first sank into the cluster center because of the dynamical friction\index{dynamical friction}
(or they were already present into the cluster core), and then have
been driven into the mass-transfer regime by hardening processes
induced by gravitational interactions during the core collapse phase.
According to this scenario, {\it the double BSS sequence detected in
  M30 dates the occurrence of the  core collapse event back to 1--2
  Gyr ago. If the proposed scenario is correct, this discovery opens
  the possibility of defining a powerful ``clock'' to date the
  occurrence of this dramatic event in a star cluster history} (see
also Section \ref{fersec:dyn_clock}).

Additional clues in favour of a different formation history for the
BSSs belonging to the two sequences are suggested by their central
concentration and specific frequency. The red-BSSs are more centrally
segregated than the blue ones (Fig.~ \ref{double_cumu}), with no
red-BSSs observed at $r>30\arcsec$ ($\sim 1.3$ pc) from the cluster
center.  Moreover, the value of the BSS specific frequency with
respect to HB stars varies significantly over the cluster extension,
reaching the surprising value of $\sim 1.55$ when only the central
cusp of the star density profile ($r<5\arcsec-6\arcsec$) is
considered.  This is the highest BSS specific frequency measured in
any GC \cite{fe03sixGCs}, and it further supports the
possibility that in M30 we are observing the effect of an enhanced
gravitational interaction activity on single and binary stars.
 
The proposed picture leads to a testable observational prediction: the
red-BSS sequence should be populated by binaries with short orbital
periods\index{short orbital period binary}. A recent paper \cite{knigge09} suggested that the dominant
BSS formation channel is the evolution of binary systems\index{binary system},
independently of the dynamical state of the parent cluster. The double
BSS sequence in M30 possibly shows that binary evolution alone does
not paint a complete picture: dynamical processes can indeed play a
major role in the formation of BSSs. Interestingly, preliminary
indications of double BSS sequences have been collected for two
additional clusters: M15\index{M15} (Beccari et al. 2014, in preparation) and
NGC\,362\index{NGC 362} \cite{dale2013}.  Moreover,
detailed spectroscopic investigations are certainly worth performing
to obtain a complete characterisation of the BSS properties (orbital
periods, rotation velocities, etc.).  In this respect particularly
promising is the search for the chemical signature of the MT process
(see Section \ref{fersec:spec}) for the BSSs along the red sequence.

\begin{figure}
\begin{center} 
\includegraphics[width=119mm]{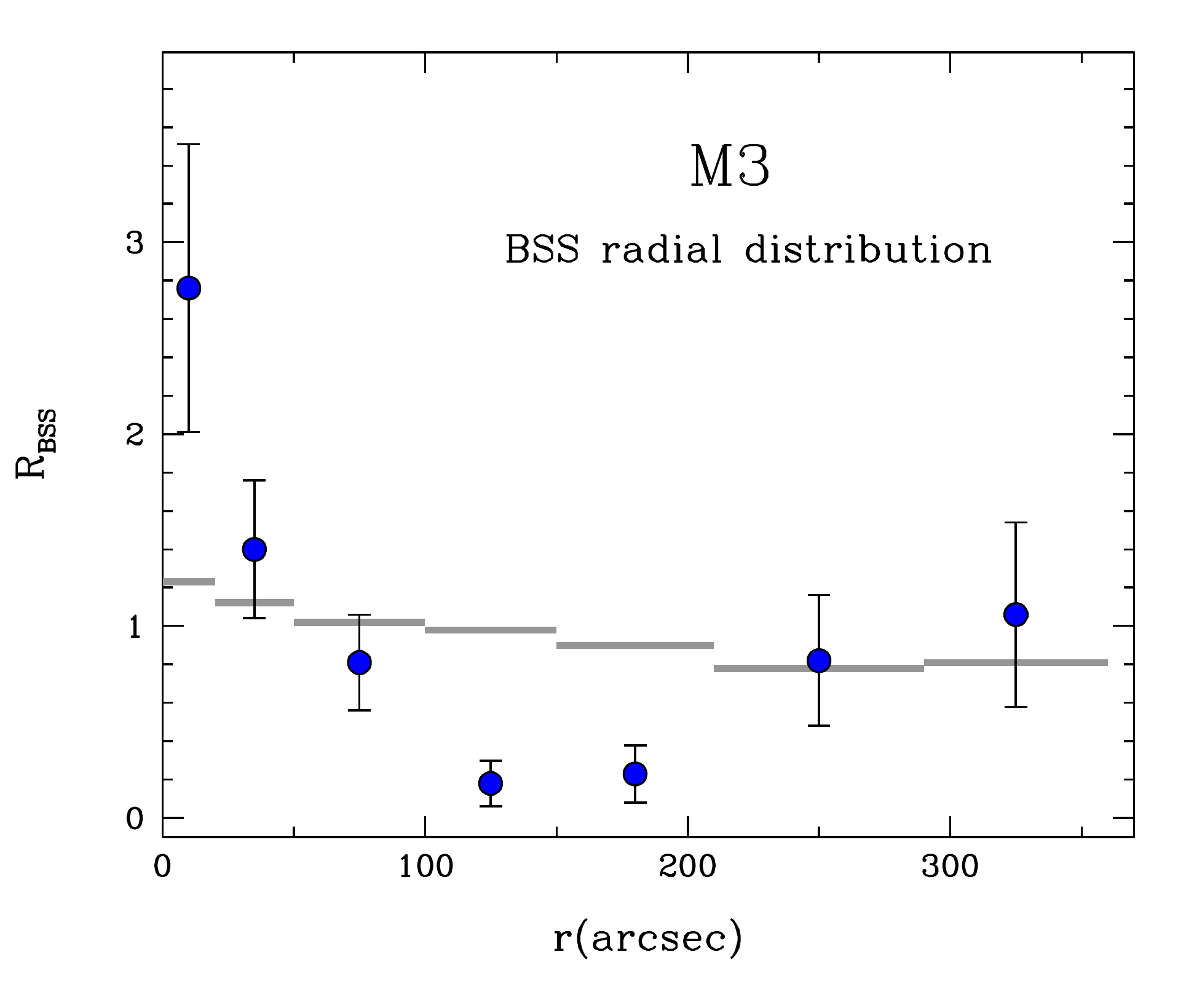}
\caption{Bimodal radial distribution of BSSs in M3\index{M3}. Blue dots mark the
  value of the BSS double normalised ratio as defined in
  eq. (\ref{eq:rbss}), computed at various distances from the cluster
  centre, the gray segments mark the double normalised ratio of HB
  stars. From \cite{fe97}.}
\label{rpop_m3}
\end{center}
\end{figure}

\section{The BSS Radial Distribution}
\label{fersec:rad_distr}
M3\index{M3} has played a fundamental role in the BSS history: not only it is
the system where BSSs have been first identified \cite{sand53}, but
also where their radial distribution has been studied over the entire
cluster extension for the first time (F97).  In fact by combining UV\index{ultraviolet}
HST\index{Hubble Space Telescope} observations of the cluster central region (F97) and wide field
ground-based observations in the visible-light bands \cite{fe93,buon94},
F97 presented the BSS radial distribution of M3 out to $r\sim 6'$. The
result was completely unexpected: BSSs appeared to be more centrally
concentrated than RGB stars in the cluster central regions, and less
concentrated in the outskirts. The result is shown in Fig.~
\ref{rpop_m3} and it clearly shows that the radial distribution of
BSSs in M3 is bimodal: it reaches a maximum in the centre of the
cluster, shows a clear-cut dip in the intermediate region (at
$100\arcsec<r< 200\arcsec$), and rises again in the outer region.
 
Sigurdsson, Daviest \& Bolte \cite{sigurd94} suggested that the bimodal BSS distribution observed
in M3 could be explained by assuming that all BSSs formed in the core
by direct collisions\index{collision} (thus creating the central peak of the
distribution) and some of them were kicked out from the centre by the
recoil of the interactions. Those BSSs ejected to a few core radii
rapidly drifted back to the centre of the cluster due to mass
segregation\index{mass segregation} (thus contributing to the central BSS concentration and
generating the paucity of BSSs at intermediate distances of a few core
radii). BSSs affected by more energetic recoils would have been kicked
out to larger distances and, since they require much more time to
drift back toward the core, they may account for the overabundance of
BSSs observed in the cluster outskirts.  However, Monte-Carlo
dynamical simulations\index{dynamical simulation} \cite{mapelli04, mapelli06} demonstrated that
BSSs kicked out from the core either are lost from the cluster, or
sink back to the centre in 1--2 Gyr only. Hence, the observed BSS
bimodal distributions cannot be explained with a purely collisional
population, and to accurately reproduce the external upturn of the
distribution it is necessary to assume a sizable fraction ($\sim
20-40\%$) of MT-BSSs, generated in the peripheral regions where
primordial binaries can evolve in isolation and experience mass
transfer processes without suffering significant interactions with
other cluster stars.
 
While the bimodality detected in M3 was considered for years to be
{\it peculiar}, more recent results demonstrated that this is not the
case.  In fact, the same observational strategy adopted by F97 in M3
has been applied to a number of other clusters, and bimodal
distributions have been detected in the majority of cases ($\sim 15$)
studied so far. Examples can be found in:  47\,Tuc\index{47 Tucanae} \cite{fe04} ,
  NGC\,6752\index{NGC 6752}\cite{sabbi04}, 
 M55\index{M55} \cite{zaggia97,lan07M55}, M5\index{M5} \cite{w06,lan07M5}, 
 NGC\,6388\index{NGC 6388} \cite{ema08_6388},  M53\index{M53} \cite{bec07}.  
 Only a few exceptions are
known: M79\index{M79} and M75\index{M75}, which do not present any external upturn
\cite{lan071904, contrerasm75}, and three clusters
($\omega$\,Centauri\index{$\omega$ Centauri}, NGC\,2419\index{NGC 2419} and Palomar 14\index{Palomar 14}) showing a completely
flat BSS radial distribution, totally consistent with that of the
reference population
(\cite{fe06,ema_2419,beccari11pal14}, respectively).  The last three
cases deserve a specific comment.  The flat behaviour discovered in
these clusters suggests that the BSS radial distribution is not yet
significantly altered by stellar interactions and by the dynamical
evolution of the cluster. {\it Indeed, this is the cleanest evidence
  of the fact that these systems are not fully relaxed yet, even in
  the central regions}.  We emphasise that this result is much more
solid than any other estimate of mass segregation in these stellar
systems.  In fact, the degree of mass segregation is usually evaluated
from star number counts along the MS, down to quite faint magnitudes
where incompleteness biases can be severe (see,
  e.g., \cite{anderson02,jordi09}). Instead the computation of the BSS
specific frequency refers to much brighter objects, as BSS and
TO/RGB/HB stars.

\begin{figure}
\begin{center} 
\includegraphics[width=119mm]{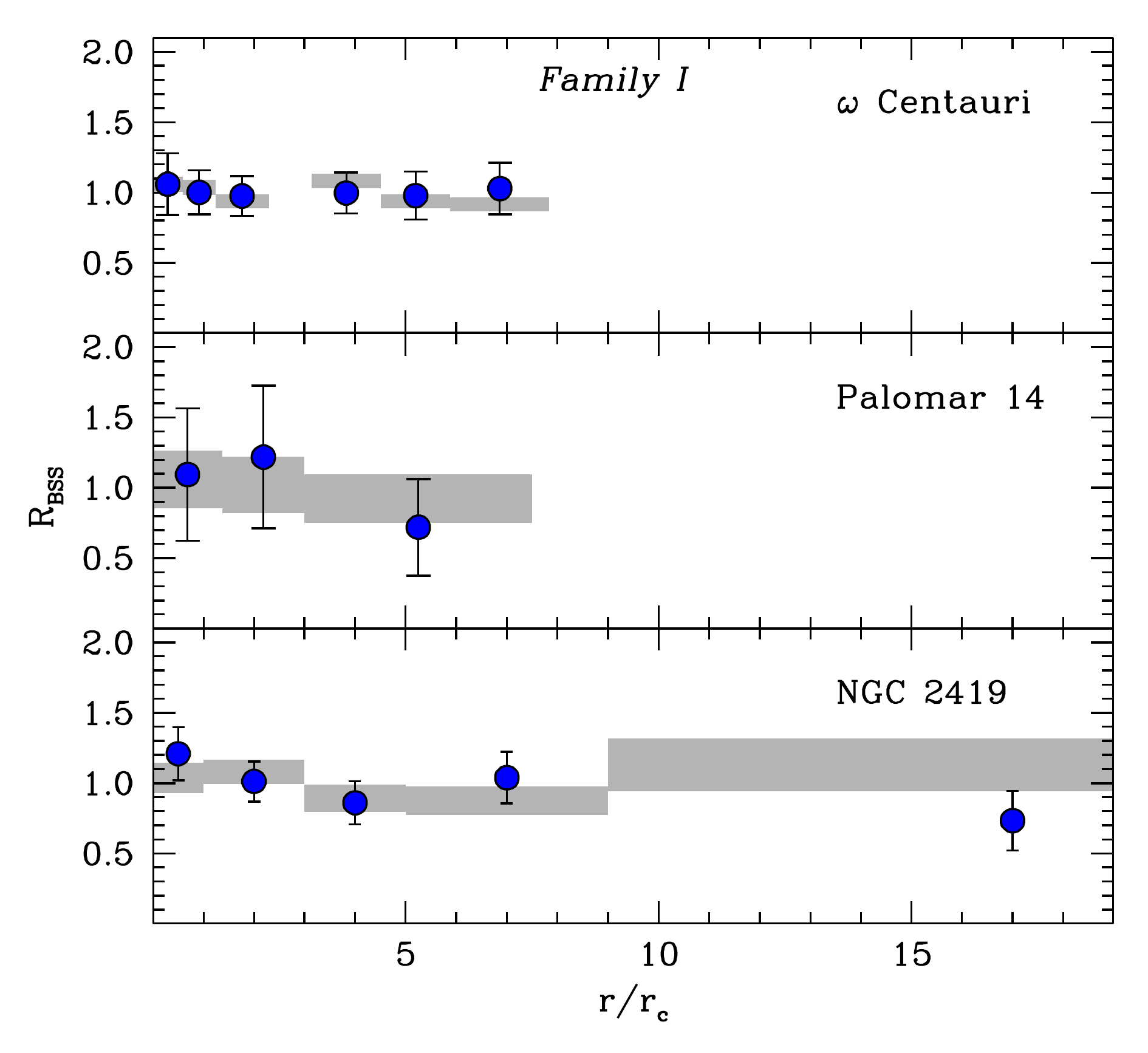}
\caption{BSS radial distribution observed in $\omega$\,Centauri,
  Palomar\,14 and NGC\,2419, with the blue circles marking the values
  of $R_{\rm BSS}$, defined in eq. (\ref{eq:rbss}). The distribution of the
  double normalised ratio measured for RGB or HB stars is also shown
  for comparison (grey strips). The BSS radial distribution is
  flat and totally consistent with that of the reference population,
  thus indicating a low degree of dynamical evolution for these three
  GCs (\emph{Family I}). From \cite{fe12}.}
\label{dyncl1}
\end{center}
\end{figure}

\begin{figure}
\begin{center} 
\includegraphics[width=119mm]{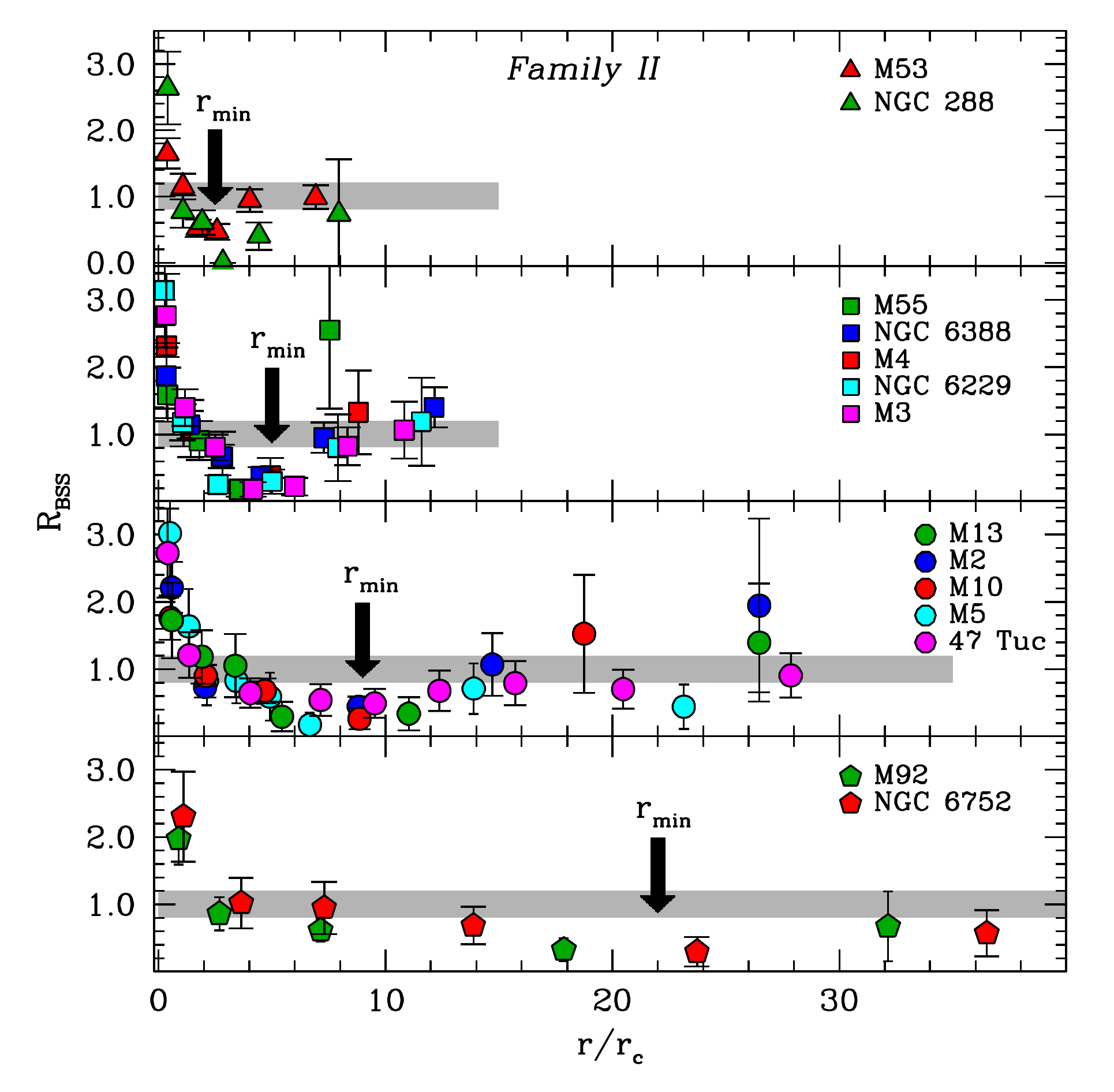}
\caption{BSS radial distribution observed in clusters of intermediate
  dynamical age (\emph{Family II}). The distribution is clearly
  bimodal and the radial position of the minimum (marked with the
  arrow and labelled as $r_{\rm min}$) clearly moves outward from top
  to bottom, suggesting that the bottom clusters are more dynamically
  evolved than the upper ones.  For the sake of clarity, the grey
  bands schematically mark the distribution of the reference
  populations. From \cite{fe12}.}
\label{dyncl2}
\end{center}
\end{figure}

\begin{figure}
\begin{center} 
\includegraphics[width=119mm]{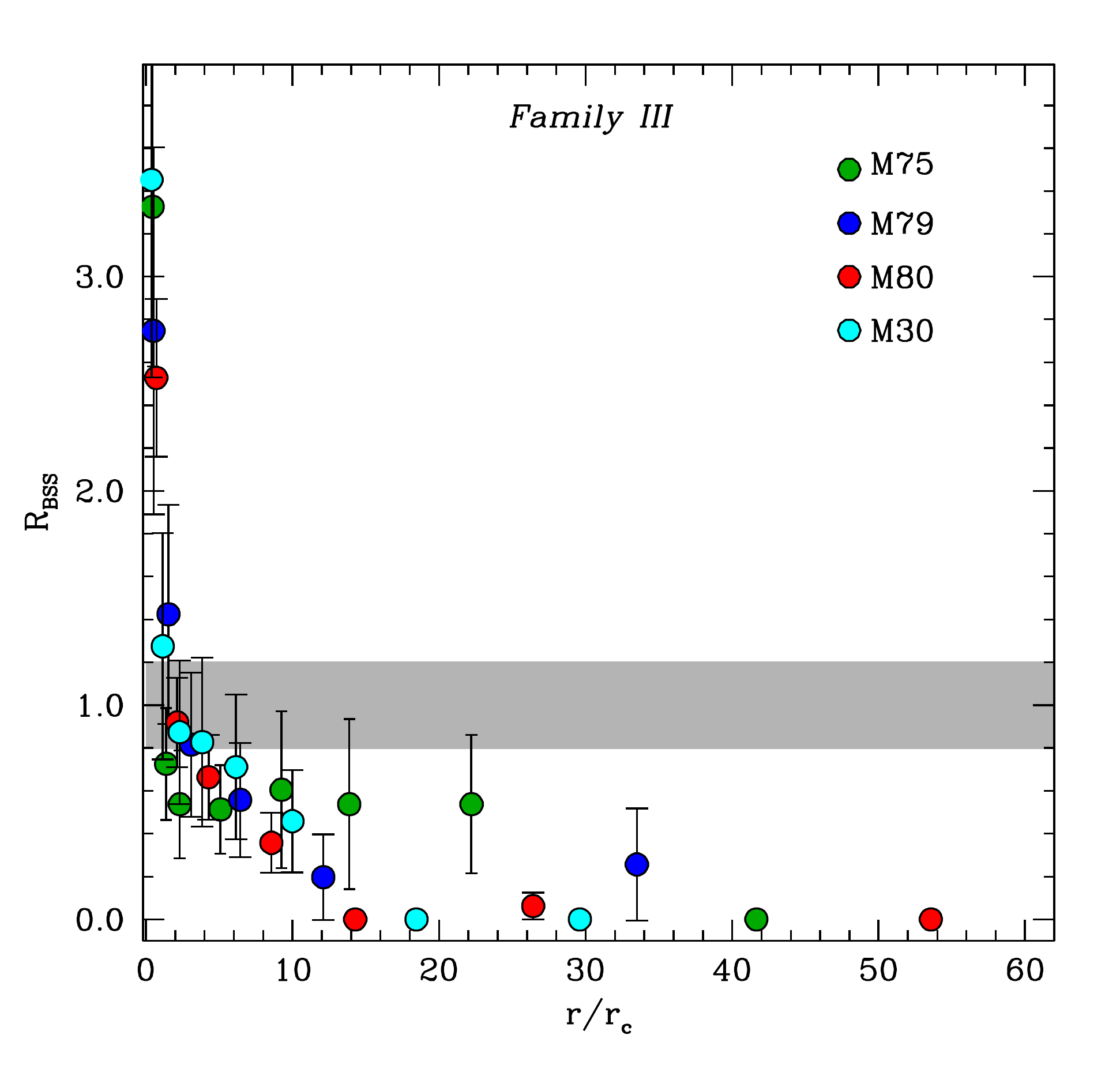}
\caption{BSS radial distribution for dynamically old clusters
  (\emph{Family III}): only a central peak is visible, while the
  external upturn is no more present because, within the proposed
  scenario, the dynamical friction\index{dynamical friction} has been efficient out to the
  cluster outskirts. From \cite{fe12}.}
\label{dyncl3}
\end{center}
\end{figure}

\section{Setting the Dynamical Clock for Stellar Systems}
\label{fersec:dyn_clock}
 
The entire database of available BSS radial distributions has been
analysed by us (\cite{fe12}, hereafter F12).  Such a dataset contains 21
GCs with very different structural properties (hence possibly at
different stages of their dynamical evolution), but with nearly the
same chronological age (12-13 Gyr; \cite{marinfranch09}), with the
only exception of Palomar 14\index{Palomar 14} which formed $\sim 10.5$ Gyr ago
\cite{dotter08}.  While significant cluster-to-cluster variations
were already known, F12 discovered that once the radial distance from
the centre is expressed in units of the core radius (thus to allow a
meaningful comparison among the clusters), GCs can be efficiently
grouped on the basis of the shape of their BSS radial distribution,
and at least three families can be defined:
\begin{itemize}
\item \emph{Family I --} the radial distribution of the BSS double
  normalised ratio ($R_{\rm BSS}$) is fully consistent with that of
  the reference population ($R_{\rm pop}$) over the entire cluster
  extension (see Fig.~ \ref{dyncl1});
\item \emph{Family II --} the distribution of $R_{\rm BSS}$ is
  incompatible with that of $R_{\rm pop}$, showing a significant
  bimodality, with a central peak and an external upturn. At
  intermediate radii a minimum is evident and its position ($r_{\rm
    min}$) can be clearly defined for each sub-group (see Fig.~
  \ref{dyncl2});
\item \emph{Family III --} the radial distribution of $R_{\rm BSS}$ is
  still incompatible with that of the reference population,
  showing a well defined central peak with no external upturn (see
  Fig.~ \ref{dyncl3}).
\end{itemize}
\emph{Which is the physical origin of these distributions?}  Previous
preliminary analysis \cite{mapelli04,mapelli06,lan07M5,lan071904} of a few clusters indicated that BSSs
generated by stellar collisions\index{collision} mainly/only contribute to the central
peak of the distribution, while the portion beyond the observed
minimum is populated by MT-BSSs which are evolving in isolation in the
cluster outskirt and have not yet suffered the effects of dynamical
friction (see Sect. \ref{fersec:rad_distr}).  Overall, the BSS radial
distribution is primarily modelled by the long-term effect of
dynamical friction\index{dynamical friction} acting on the cluster binary population (and its
progeny) since the early stages of cluster evolution. In fact, what we
call MT-BSS today is the by-product of the evolution of a $\sim 1.2
$M$_\odot$ binary that has been orbiting the cluster and suffering the
effects of dynamical friction for a significant fraction of the
cluster lifetime.  The efficiency of dynamical friction decreases for
increasing radial distance from the centre, as a function of the local
velocity dispersion and mass density.  Hence, dynamical friction first
segregates (heavy) objects orbiting close to the centre and produces a
central peak in their radial distribution. As the time goes on, the
effect extends to larger and larger distances, thus yielding to a
region devoid of these stars (i.e., a dip in their radial
distribution) that progressively propagates outward.  Simple
analytical estimate of the radial position of this dip turned out to
be in excellent agreement with the position of the minimum in the
\emph{observed} BSS radial distributions ($r_{\rm min}$), despite a
number of crude approximation (see, e.g., \cite{mapelli06}).
Moreover, a progressive outward drift of $r_{\rm min}$ as a function
of time is confirmed by the results of direct N-body simulations that
follow the evolution of $\sim 1.2$ M$_\odot$ objects within a reference
cluster over a significant fraction of its lifetime.

In light of these considerations, the three families defined in
Figs. \ref{dyncl1}--\ref{dyncl3} correspond to GCs of increasing
dynamical ages\index{dynamical age}. Hence, the shape of the BSS radial distribution turns
out to be a powerful dynamical-age indicator. A flat BSS radial
distribution (consistent with that of the reference population; see
{\it Family I} in Fig. \ref{dyncl1}) indicates that dynamical friction
has not played a major role yet even in the innermost regions, and the
cluster is still dynamically young. This interpretation is confirmed
by the absence of statistically significant dips in the BSS
distributions observed in dwarf spheroidal galaxies (\cite{mapelli09, monelli12}; see also Chap. 6): these
are, in fact, collisionless systems where dynamical friction is
expected to be highly inefficient.  In more evolved clusters ({\it
  Family II}), dynamical friction starts to be effective and to
segregate BSSs that are orbiting at distances still relatively close
to the centre: as a consequence, a peak in the centre and a minimum at
small radii appear in the distribution, while the most remote BSSs are
not yet affected by the action of dynamical friction (this generates
the rising branch of the observed bimodal BSS distributions; see upper
panel in Fig. \ref{dyncl2}).  Since the action of dynamical friction
progressively extends to larger and larger distances from the centre,
the dip of the distribution progressively moves outward (as seen in
the different groups of {\it Family II} clusters; Fig. \ref{dyncl2},
panels from top to bottom).  In highly evolved systems dynamical
friction already affected even the most remote BSSs, which started to
gradually drift toward the centre: as a consequence, the external
rising branch of the radial distribution disappears (as observed for
{\it Family III} clusters in Fig. \ref{dyncl3}). All GCs with a
single-peak BSS distribution can therefore be classified as
``dynamically old''.

Interestingly, this latter class includes M30\index{M30} (see Section
\ref{fersec:double}), a system that already suffered core collapse which
is considered as a typical symptom of extreme dynamical evolution
\cite{meylanheggie97}.  The proposed classification is also
able to shed light on a number of controversial cases debated into the
literature.  In fact, M4\index{M4} turns out to have an intermediate dynamical
age, at odds with previous studies suggesting that it might be in a
PCC state \cite{heggie08m4}. On the other hand, NGC 6752\index{NGC 6752} turns out
to be in a quite advanced state of dynamical evolution, in agreement
with its observed double King profile indicating that the cluster core
is detaching from the rest of the cluster structure \cite{fe036752}.
Finally this approach might provide the key to discriminate between a
central density cusp due to core collapse (as for M30) and that due to
the presence of an exceptional concentration of dark massive objects
(neutron stars\index{neutron star} and/or the still elusive intermediate-mass black
  holes\index{black hole}; see the case of NGC 6388\index{NGC 6388} in \cite{lan07imbh,lan13}, and references
  therein).

\begin{figure}
\begin{center} 
\includegraphics[width=119mm]{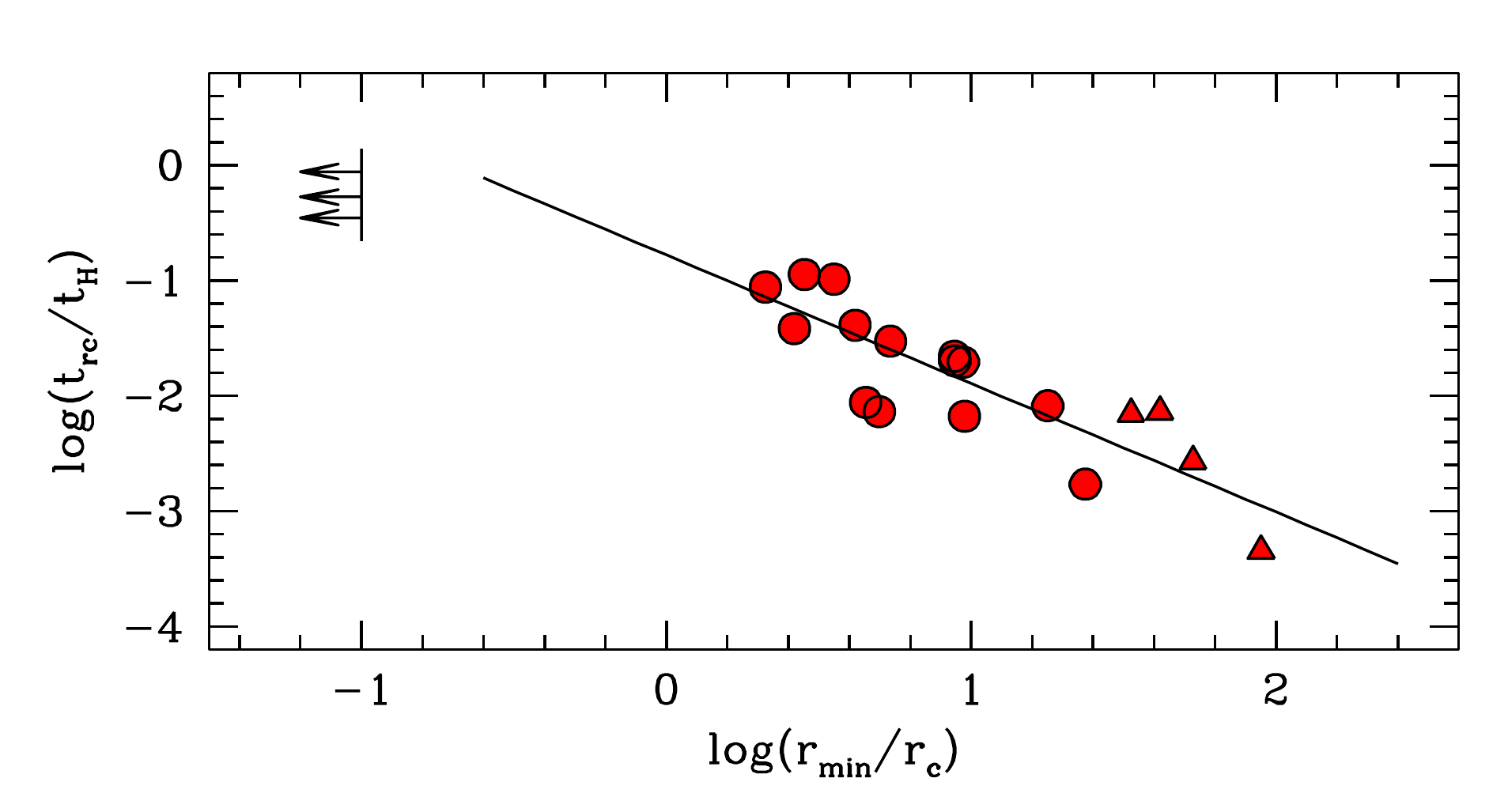}
\caption{Core relaxation time (normalised to the Hubble time\index{Hubble time} $t_H$) as
  a function of the clock hand of the proposed \emph{dynamical clock}
  ($r_{\rm min}$, in units of the core radius). Dynamically young
  systems (\emph{Family I}) show no minimum and are plotted as
  lower-limit arrows at $r_{\rm min}/r_c=0.1$.  For dynamically old
  clusters (\emph{Family III}, triangles), the distance of the
  farthest radial bin where no BSSs are observed has been adopted as
  $r_{\rm min}$. As expected for a meaningful clock, a tight
  anticorrelation is found: clusters with relaxation times of the
  order of the age of the Universe show no signs of BSS segregation
  (hence their BSS radial distribution is flat and $r_{\rm min}$ is
  not definable; see Fig. \ref{dyncl1}), whereas for decreasing
  relaxation times the radial position of the minimum increases
  progressively.  The solid line correspond to the best-fit relations,
  given in (\ref{eq:dyncl_trc}).  From \cite{fe12}.}
\label{dyncl4}
\end{center}
\end{figure}

The quantisation in distinct age-families is of course an
over-simplification, while the position of $r_{\rm min}$ is found to
vary with continuity as a sort of clock time-hand.  This allowed F12
to define the first empirical clock able to measure the dynamical age
of a stellar system from pure observational quantities (the {\it
 dynamical clock}): as the engine of a chronometer advances the
clock hand to measure the time flow, in a similar way the progressive
sedimentation of BSSs towards the cluster centre moves $r_{\rm min}$
outward, thus marking its dynamical age.  This is indeed confirmed by
the tight correlations found between the clock-hand ($r_{\rm min}$)
and the central and half-mass relaxation times ($t_{\rm rc}$ and
$t_{\rm rh}$, respectively), which are commonly used to measure the
cluster dynamical evolution\index{dynamical evolution} time-scales. The trend with $t_{\rm rc}$
found by F12 is shown in Fig.~ \ref{dyncl4} and the best-fit
relations is:
\begin{equation}
 \log (t_{\rm rc}/t_{\rm H}) = -1.11 \times \log(r_{\rm min}) -0.78
\label{eq:dyncl_trc}
\end{equation}
where $t_{\rm H}$ is the Hubble time.  Note that, while $t_{\rm rc}$
and $t_{\rm rh}$ provide an indication of the relaxation timescales at
specific radial distances from the cluster centre ($r_{\rm c}$ and
$r_{\rm h}$, respectively), the dynamical clock here defined provides
a measure of the global dynamical evolution of the systems, because
the BSS radial distribution simultaneously probes all distances from
the cluster centre.

\begin{figure}
\begin{center} 
\includegraphics[width=119mm]{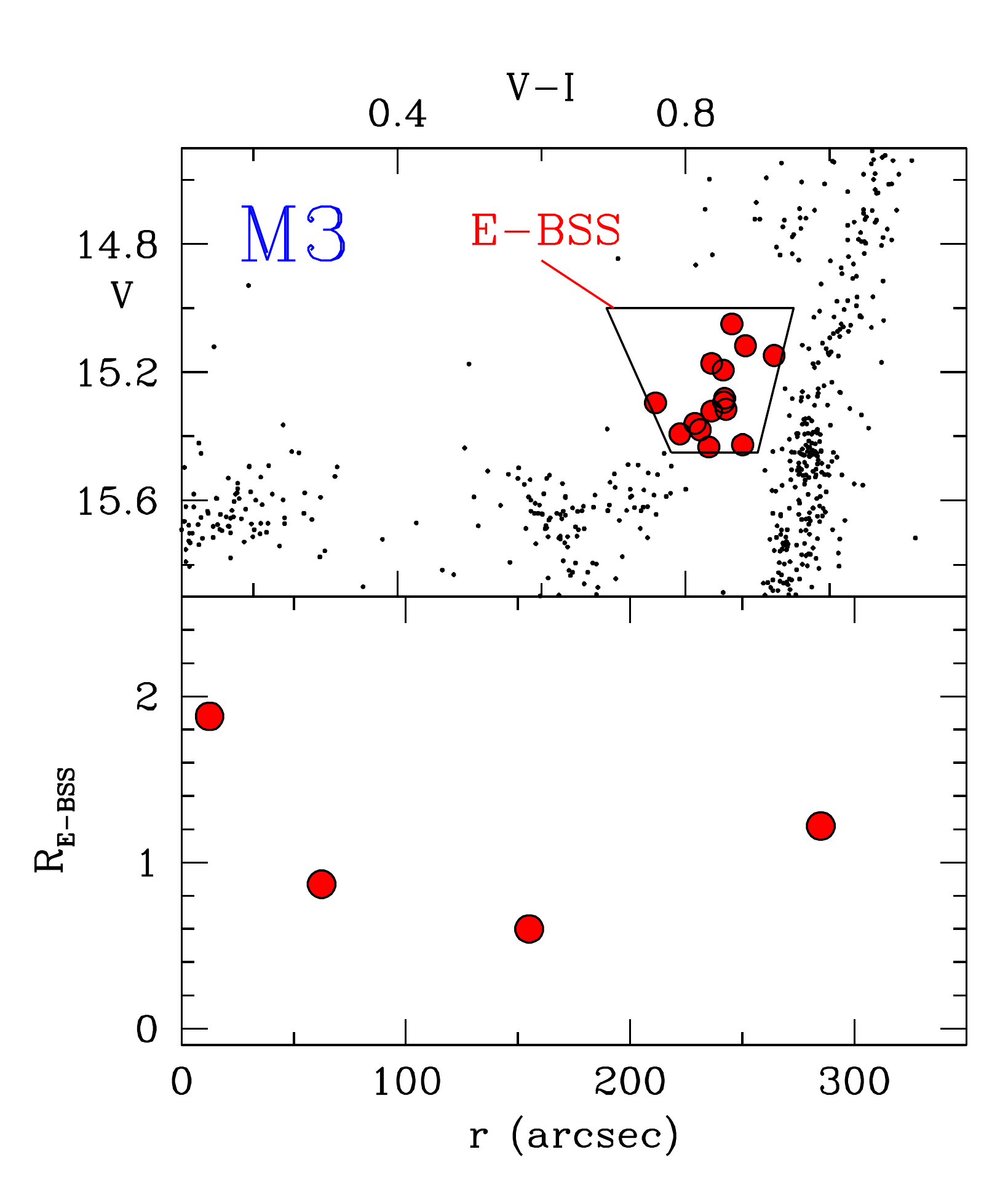}
\caption{\emph{Upper panel}: CMD of M3 zoomed in the HB/AGB
  region. Red circles and the box mark the sample of candidate evolved
  BSSs. \emph{Lower panel}: Double normalised ratio computed for the
  sample of candidate E-BSSs. A bimodality similar to that found for
  BSSs (see Fig. \ref{rpop_m3}) is clearly visible. From
  \cite{fe97}.}
\label{ebss_m3}
\end{center}
\end{figure}

\begin{figure}
\begin{center} 
\includegraphics[width=119mm]{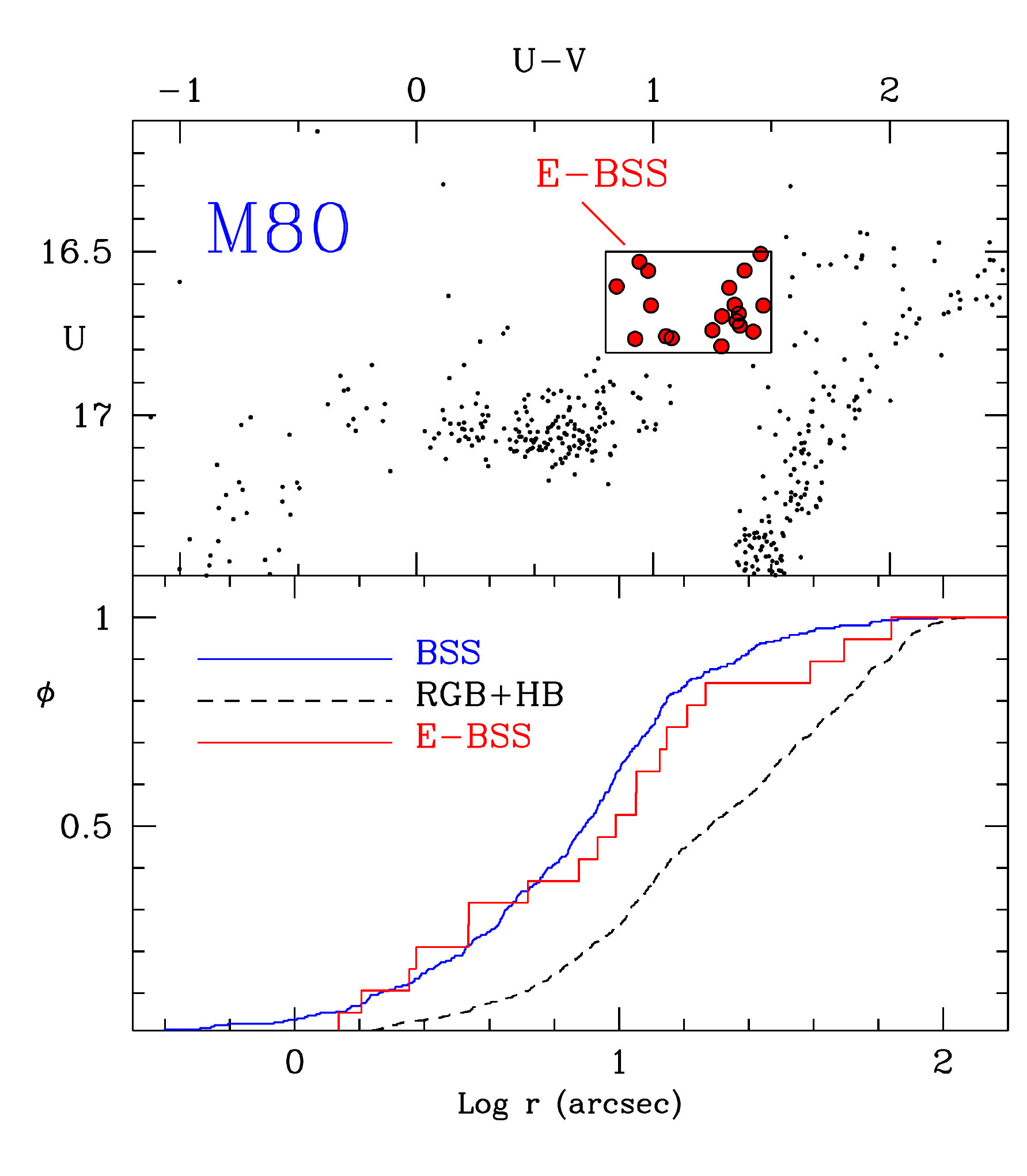}
\caption{\emph{Upper panel}: CMD of M80 zoomed in the HB/AGB
  region. Red circles and the box mark the sample of candidate evolved
  BSSs. \emph{Lower panel}: Cumulative radial distribution of BSSs
  (blue line), RGB+HB stars (black dashed line) and candidate evolved
  BSSs (red line). Clearly, E-BSSs share the same radial distribution
  of BSSs and are significantly more segregated than normal cluster
  stars (RGB and HB stars). From \cite{fe99}.}
\label{ebss_m80}
\end{center}
\end{figure}

\begin{figure}
\begin{center} 
\includegraphics[width=119mm]{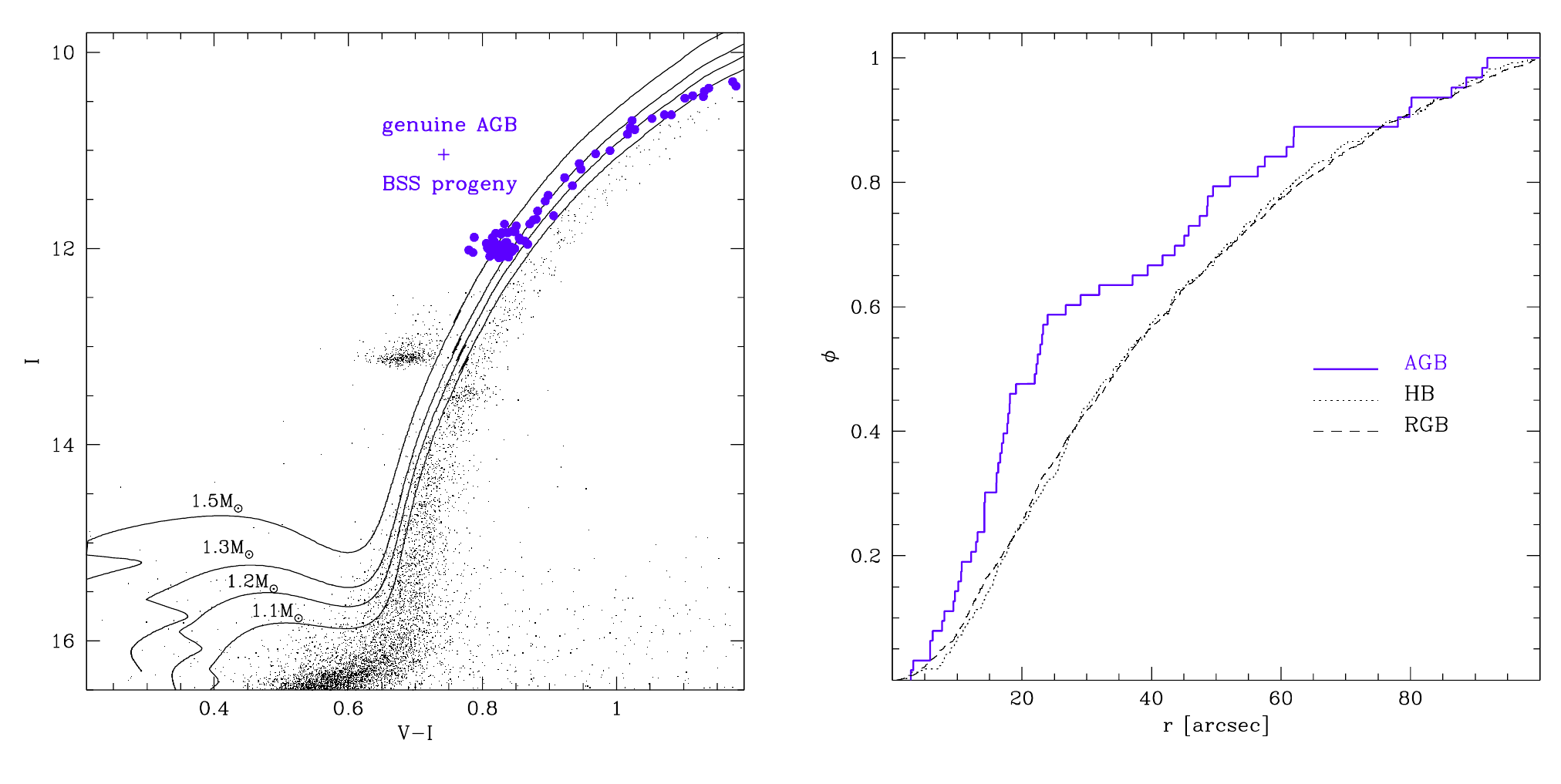}
\caption{\emph{Left panel}: Portion of the CMD of 47\,Tucanae\index{47 Tucanae} above
  the MS-TO.  Blue circles highlight the AGB population, which is
  suspected to be severely contaminated by a sample of E-BSSs. Solid
  lines are theoretical tracks from Pietrinferni et al. \cite{pietrinferni06} for
  stars with masses ranging between 1.1 M$_\odot$ and 1.5 M$_\odot$
  (see labels), showing that the RGB phase of these massive objects is
  well superposed to the AGB sequence of the cluster. \emph{Right
    panel}: Cumulative radial distribution of the AGB population,
  likely contaminated by E-BSSs (blue solid line), compared to that of
  HB stars (black dotted line) and RGB stars (black dashed
  line). Clearly, the ``AGB''\index{asymptotic giant branch} population is significantly more
  segregated than normal cluster stars, as expected if it is
  contaminated by more massive objects, as E-BSSs. From
  \cite{bec06}.}
\label{ebss_47tuc}
\end{center}
\end{figure}

\section{Searching for the BSS Progeny:  Evolved BSSs}
\label{fersec:ebss}

\emph{Although BSSs have been routinely observed for 60 years now, no
  firm identification of even a single evolved BSS (E-BSS) has been
  obtained to date.}  Indirect evidence of the possible existence of
E-BSSs has been derived from photometric\index{photometry} criteria. 
Renzini \& Fusi
  Pecci \cite{renzfusi88}
suggested to search for E-BSSs during their core helium burning phase,
when they should appear redder and brighter than {\it normal} HB\index{horizontal branch star}
stars, i.e. they should be located in a region of the CMD between the
HB level and the Asymptotic Giant Branch (AGB).  Following this
prescription Fusi Pecci and collaborators \cite{fp92} identified a few E-BSS candidates in several
clusters with predominantly blue HBs, where the likelihood of
confusing E-BSSs with true HB or evolved HB stars was minimised.
Following the same prescription, F97 identified a sample of E-BSS
candidates in M3\index{M3} (see upper panel in Fig.~ \ref{ebss_m3}),
demonstrating that their radial distribution is similar to that
observed for BSSs (see lower panel in Fig. \ref{ebss_m3} and
Fig. \ref{rpop_m3}).  Similar results have been obtained by us
\cite{fe99} in the case of M80\index{M80} (Fig.~ \ref{ebss_m80}).  The
cumulative distribution of E-BSSs is consistent with that of BSSs and
significantly different from that of genuine HB+RGB stars. Indeed, the
Kolmogorov-Smirnov probability that E-BSSs and BSSs are extracted from
the same parent population is $\sim 67\%$, while the same probability
between E-BSSs and RGB stars decreases to only $\sim 1.6 \%$.  This
result confirms the expectation that E-BSSs share the same radial
distribution of BSSs, both being more massive than the bulk of the
cluster stars.  It is interesting to note that the ratio between the
number of bright BSSs and that of E-BSSs is $N_{\rm bBSS}/N_{\rm EBSS}
\approx 6.5$ in both GCs \cite{fe97,fe99}.  An approximate
estimate of the lifetime ratio between BSSs (i.e., BSSs in the MS
evolutionary stage) and their progeny can be obtained from the ratio
between the total number of BSSs and that of E-BSSs.  This turns out
to range between 11 and 16, thus suggesting that we should expect 1
E-BSS ever 13 genuine BSSs.  Indeed this is in very good agreement
with the predictions of recent theoretical models of E-BSSs
\cite{sills09}.
 
More recently, we \cite{bec06} discovered a very promising signature of
the existence of BSS descendants along the AGB of 47\,Tuc\index{47 Tucanae}: a
\emph{significant excess} of stars which are \emph{more centrally
  segregated} than the RGB and HB populations has been found in the
AGB region of the CMD (Fig.~ \ref{ebss_47tuc}).  Within $1'$ from the
cluster centre $\sim 53$ ``AGB stars''\index{asymptotic giant branch star} are counted, while only $\sim
38$ such objects are predicted on the basis of the HB star number
counts and the post-MS evolutionary timescales
\cite{renzfusi88}: this makes an excess of $\sim 40\%$!
Because of the typical low stellar mass along the AGB ($M\sim 0.6$
M$_\odot$), this feature is hardly understandable in terms of a mass
segregation effect on these stars.  Instead it is very likely due to a
sample of more massive objects, that, given the large population of
BSSs in 47\,Tuc, most probably are the BSS descendants.  Indeed, the
comparison with theoretical tracks \cite{pietrinferni06} and
collisional models \cite{sills09} shows that the AGB population of
47\,Tuc can be significantly contaminated by stars with masses typical
of BSSs (between $\sim 1.1$ and $\sim 1.6$ M$_\odot$), which are
currently experiencing the first ascending RGB\index{red giant}
(Fig. \ref{ebss_47tuc}).

These photometric indications are very promising.  Ongoing
spectroscopic follow-ups of E-BSS candidates selected in these three
clusters will allow the first clear cut detection of the BSS progeny.
Indeed due to their higher mass, E-BSSs are expected to be
distinguishable from genuine AGB stars on the basis of their higher
values of the surface gravity. Also their rotational and chemical
composition will be measured, thus providing the first ever collected
information about the physical properties of these peculiar stars
during evolutionary paths subsequent to the core hydrogen burning
phase.  In turn, these are new, precious ingredients for the current
and future theoretical modeling of these exotica.

\section{Chemical and Kinematical Properties of BSSs}
\label{fersec:spec}

Theoretical models still predict conflicting results on the expected
properties of BSSs generated by different production channels.  In
fact, high rotational velocities are expected for both MT-BSSs
\cite{sarna96} and COL-BSSs \cite{benz87}, but braking mechanisms
like magnetic braking\index{magnetic braking} or disk locking have been suggested to
subsequently slow down the stars, with timescales and efficiencies
which are still unknown \cite{leoliv95, sills05}.  Concerning the
chemical surface abundances, hydrodynamic simulations\index{hydrodynamic simulation} \cite{lomb95}
have shown that very little mixing is expected to occur between the
inner cores and the outer envelopes of the colliding stars. On the
other hand, signatures of mixing with incomplete CN-burning products
are expected at the surface of BSSs formed via the MT channel, since
the gas at the BSS surface is expected to come from deep regions of
the donor star, where the CNO burning occurred \cite{sarna96}.
 
\begin{figure}
\begin{center} 
\includegraphics[width=119mm]{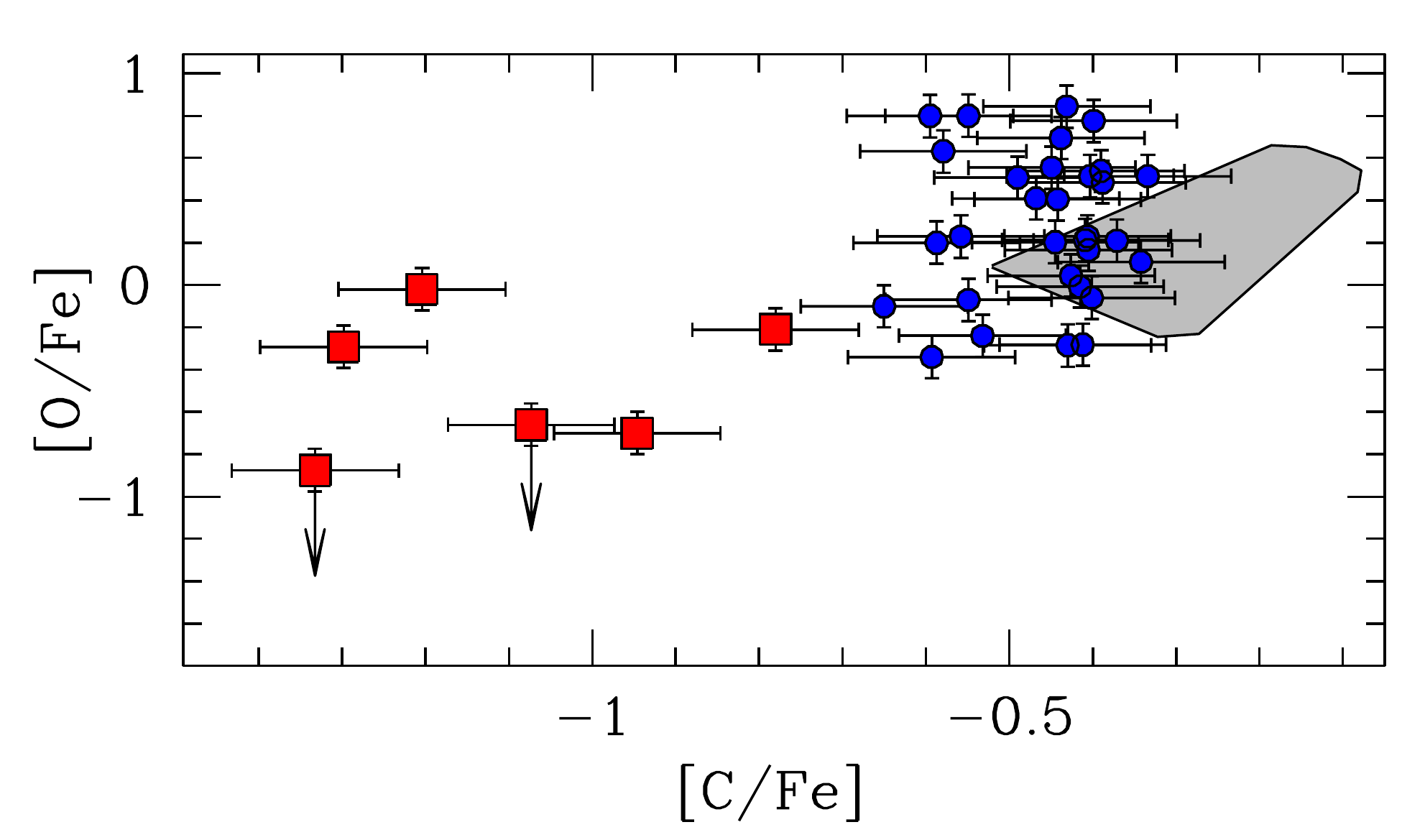}
\caption{[O/Fe] ratio as a function of [C/Fe] for the sample of 43
  BSSs observed in 47\,Tuc. Normal BSSs are marked with blue circles,
  while CO-depleted BSSs are marked with filled red squares. The gray
  region corresponds to the location of the 12 TO stars analyzed by
  \cite{carretta05}. From \cite{fe06COdep}.}
\label{COdepl}
\end{center}
\end{figure}

Sparse spectroscopic observations provided the first set of physical
properties of BSSs (effective temperature, mass, rotation
  velocity, etc.; see \cite{demarco05}), but a systematic survey of basic
parameters and surface abundance patterns was lacking, particularly in
GCs. Recently, extensive campaigns with multiplexing spectrographs
mounted at 8-m class telescopes\index{telescope} (as \emph{FLAMES}\index{FLAMES} at the\emph{ ESO-VLT}\index{Very Large Telescope}) allowed to
measure the chemical and kinematic properties for representative
numbers of BSSs in a sample of Galactic GCs.  The selected clusters
differ in dynamical state, metallicity and density, thus providing an
ideal sample for testing any possible link between the properties of
BSSs and those of the host cluster.  Indeed these observations
represent a gold mine of information, providing the first
characterisation of the structural properties of BSSs in GCs.
 
\begin{figure}
\begin{center} 
\includegraphics[width=119mm]{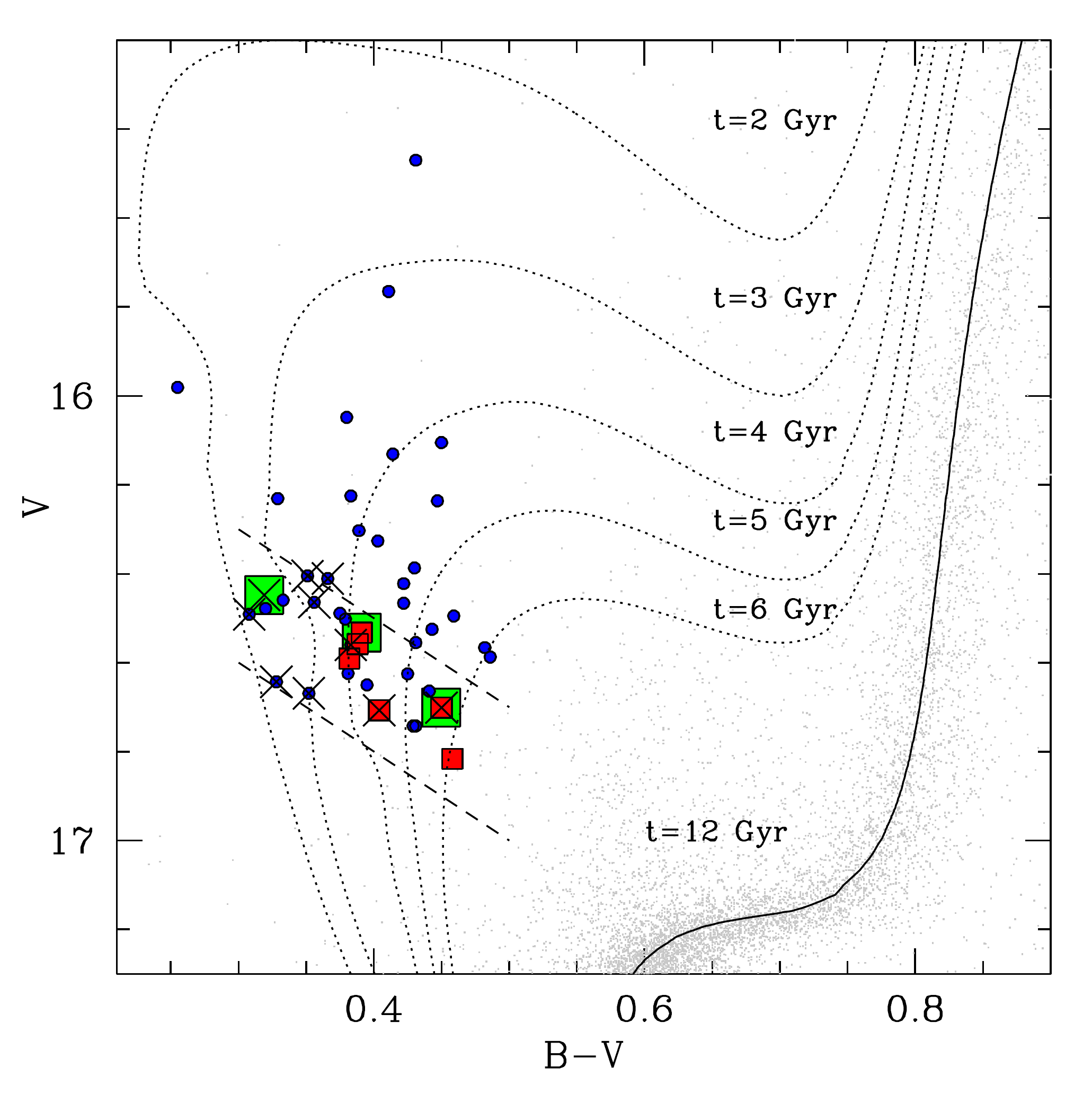}
\caption{CMD of 47\,Tuc zoomed in the BSS region.  BSSs
  showing no chemical anomalies are marked with blue circles, while
  CO-depleted BSSs are shown as red squares. Isochrones of different
  ages (from 2 to 12 Gyr) from \cite{cariulo03} are overplotted for
  comparison. The three W\,UMa systems\index{W UMa star} and the 10 BSSs rotating with
  $v \sin{i} > 10$ km/s are highlighted with large green squares and
  large crosses, respectively.}
\label{cmd_47tuc}
\end{center}
\end{figure}

The first results of this search have lead to an exciting discovery:
by measuring the surface abundance patterns of 43 BSSs in 47\,Tuc,
we \cite{fe06COdep} discovered a sub-population of BSSs with a
significant depletion of carbon\index{carbon} and oxygen\index{oxygen}, with respect to the
dominant population (see Fig.~ \ref{COdepl}).  This evidence is
interpreted as the presence of CNO burning\index{CNO burning} products on the BSS
surface, coming from the core of a deeply peeled parent star, as
expected in the case of the MT formation channel.  Thus, such a
discovery in 47\,Tuc could be the first detection of a chemical
signature clearly pointing to the MT\index{mass transfer} formation process for BSSs in a
GC.  Moreover, these observations have shown that (1) most of the BSSs
are slow rotators; (2) the CO-depleted BSSs and the few BSSs with
$v\sin i>10$ km/s appear to be ``less evolved'' than the others: they
all lie within a narrow strip at the faint-end of the BSS luminosity
distribution in the CMD (see Fig.~\ref{cmd_47tuc}); (3) some of
them are W\,UMa binaries, further suggesting that the evolution of
these systems could be a viable channel for the formation of BSSs in
GCs.

\begin{figure}
\begin{center} 
\includegraphics[width=119mm]{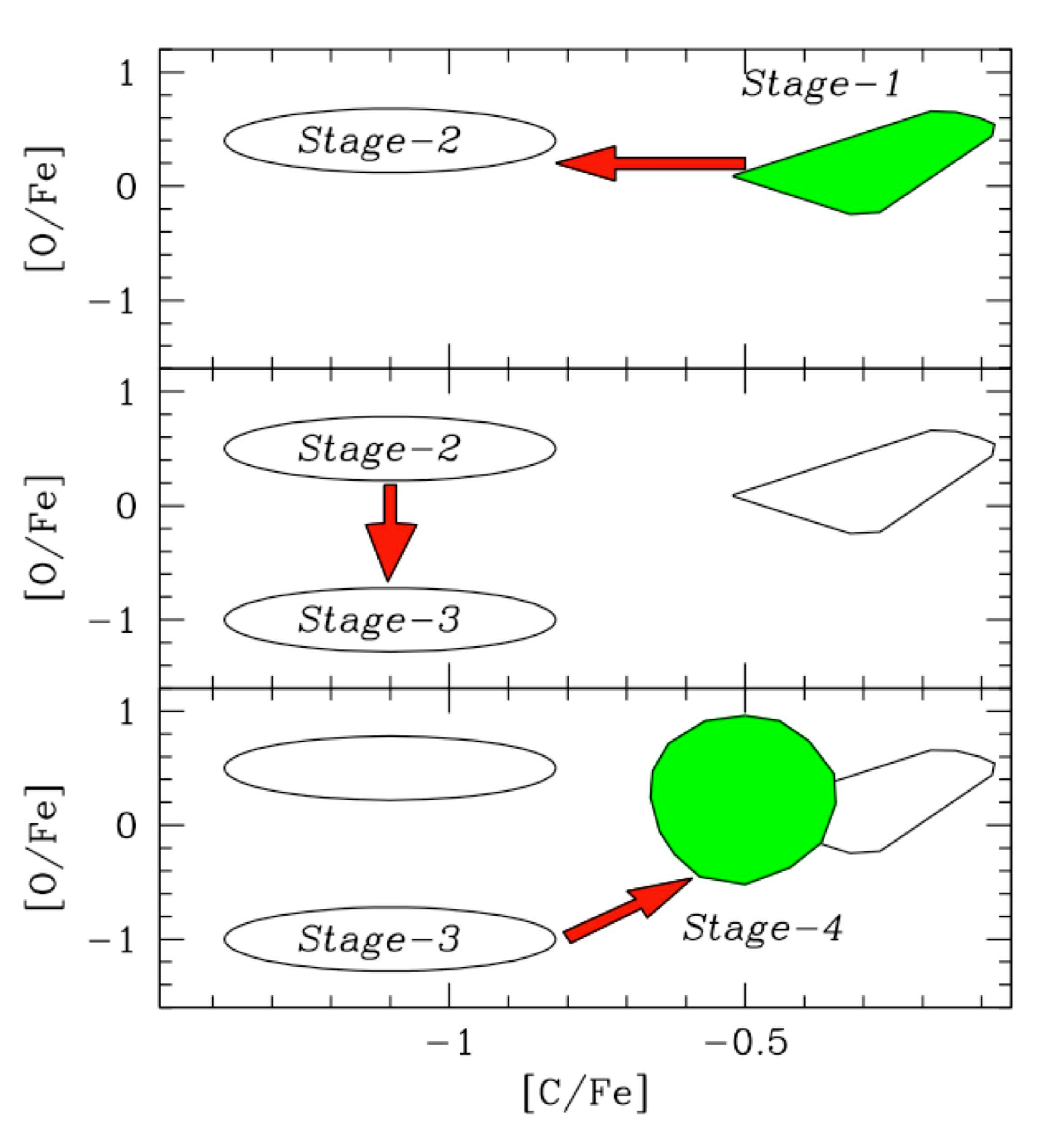}
\caption{Suggested evolution of BSSs in the [O/Fe]-[C/Fe]
  diagram. The four different stages discussed in the text are shown.}
\label{stages}
\end{center}
\end{figure}

{\it Which is the scenario emerging from these observations?}  In the
early stage of mass transfer in W\,UMa systems ({\it Stage-1}),
unprocessed material could be transferred and the resulting star would
have normal C-O abundances (see Fig.~ \ref{stages}).  As the transfer
continues, the accreted material could come from regions processed by
the CNO cycle. Hence, first C and then both C and O would appear
depleted ({\it Stage-2}). Thus it is possible to find depleted C,
normal O BSSs/W\,UMa stars. After the merger, the star would appear as
a CO-depleted non-variable BSS ({\it Stage-3}). In the sample studied
in 47\,Tuc, two or three BSSs are found in {\it Stage-2}, and 4 in
{\it Stage-3}.  The number of BSSs with CO depletion and the presence
of W\,UMa systems show that the MT channel is active even in a
high-density cluster like 47\,Tuc: at least 15\% of the BSSs are being
produced by MT.
 
The distribution of rotational velocities provide additional clues to
this scenario.  In fact, most BSSs in the 47\,Tuc sample are slow
rotators \cite{fe06COdep}, with velocities compatible with those
measured in unperturbed TO\index{turn-off} stars \cite{luc03}.  In particular, among
the three BSSs identified as W\,UMa systems, one is found to be a
rapid rotator and two are intermediate-slow rotators. This seems at
odds with what is expected, especially for W\,UMa systems which are
predicted to be rapid rotators, but, of course, different inclination
angles may play a role.  In any case, from their location in the CMD,
all the fastest BSSs are presumedly the most recently born (see
Fig. \ref{cmd_47tuc}). This is also the region of C-O depletion and
the W\,UMa behavior. The cooler, older BSSs rotate more slowly and
have ``normal'' C-O abundances.  Hence, this might suggest that during
the evolution some mixing occurred and the rotation slows down. While
rotational mixing ordinarily increases CNO anomalies, in MT-BSSs the
C-O depleted material overlies the material with normal abundances and
chemical anomalies are therefore reduced. C and possibly O would still
be low, but less so than a MT-BSS at birth. Indeed, the bulk of the
47\,Tuc sample has C roughly one half of that of the TO stars
(however, this could be due to systematics, since C in TO stars has
been measured from different lines with respect to BSSs). Future
observations will hopefully clarify this issue.

\begin{figure}
\begin{center} 
\includegraphics[width=119mm]{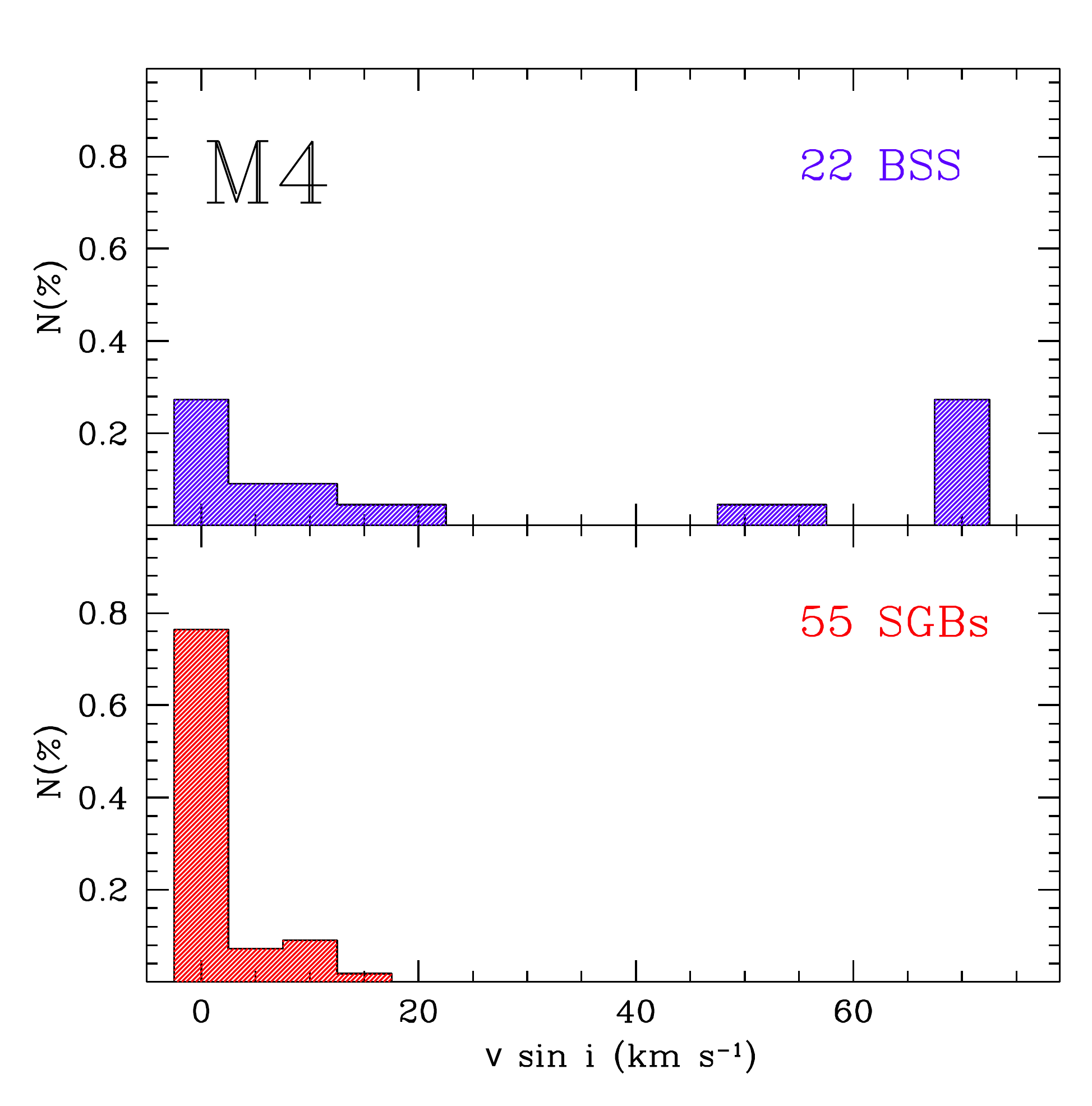}
\caption{Distribution of rotational velocities measured for 22 BSSs
  (upper panel) and 55 SGB stars (lower panel) in the GC M4. Six BSSs
  have $v\sin i\ge 70$\,km/s and are all plotted in a single
  bin. Eight BSSs have been classified as fast rotators (i.e., with
  $v\sin i>50$\,km/s) and represent the largest sample (40\% of the
  total) of rapidly rotating BSSs ever measured in a GC. From
  \cite{lovisi10}.}
\label{vrot_m4}
\end{center}
\end{figure}

\begin{figure}
\begin{center} 
\includegraphics[width=119mm]{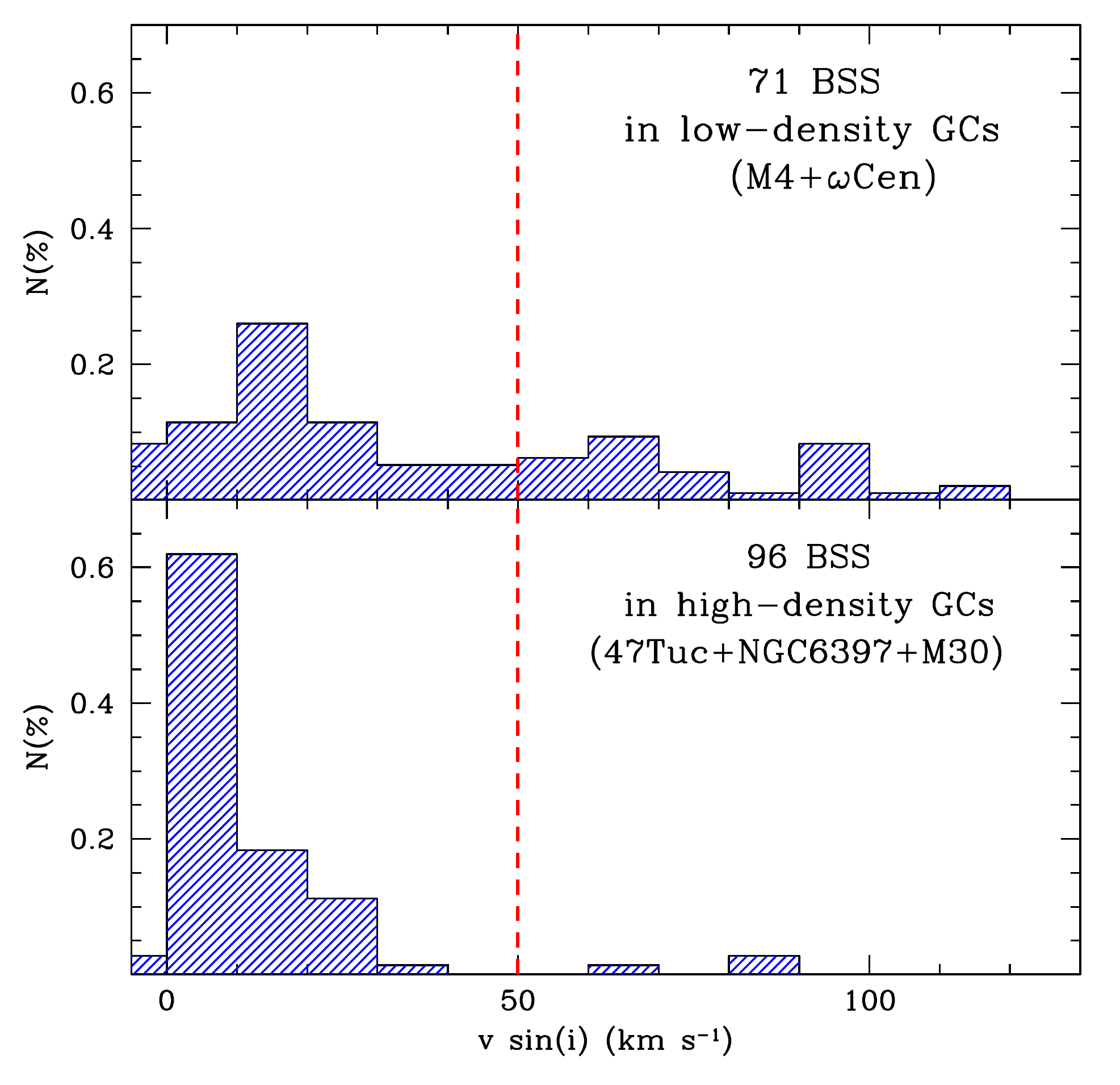}
\caption{Distribution of rotational velocities of 71 BSSs observed in
  low-density GCs (upper panel) and 96 BSSs in high-density GCs (lower
  panel). A systematic difference is apparent, possibly suggesting
  that some braking mechanism\index{braking mechanism} is active in high-density environments,
  while rapid rotation\index{rotation} can last longer in low-density
  clusters. Adapted from \cite{lovisi10, lovisi12}.}
\label{vrot_all}
\end{center}
\end{figure}

A similar study has now been extended to other clusters and up to now
more than 160 BSSs in 5 GCs (namely 47\,Tuc\index{47 Tucanae}, M4\index{M4}, NGC 6397\index{NGC 6397}, M30\index{M30} and
$\omega$Cen\index{$\omega$ Centauri}) have been observed (\cite{fe06COdep,lovisi10,lovisi12}, Mucciarelli et al. 2013 in preparation).  Unfortunately,
however, these observations provided the first observational evidence
that radiative levitation\index{radiative levitation} affects not only HB stars hotter than
11,000\,K \cite{quievy09}, but also BSSs hotter than $T>8,000$\,K
\cite{lovisi12}. The effect is clearly traced by the value of the
metallicity measured on the surface of the hottest BSSs, which is
systematically and significantly larger than that of the parent
cluster \cite{lovisi12}. Of course, in presence of radiative
levitation the measured chemical abundances cannot be interpreted in
the context of the BSS formation channels and the occurrence of this
process in the hottest (and brightest) BSSs has {\it de facto}
hampered the possibility of putting the result found in 47\,Tuc onto a
more solid statistical base.  In fact the observable pool of ``safe
BSSs'' (i.e. those cooler than 8000\,K) is quite limited in number, or
too faint for the capabilities of the current generation of
high-resolution spectrographs ($V\lsim 18-18.5$).  The results
obtained from the few BSSs not affected by radiative levitation are
the following: no chemical anomalies have been found in the sample of
11 BSSs measured in M4 \cite{lovisi10}, only one depleted BSS has
been found in M30 (Lovisi et al. 2013, ApJ submitted), and possibly
one or two have been observed in $\omega$\,Centauri (Mucciarelli et
al. 2013, in prep.).  Within the limitations of small number statistics, the
collected data confirm that the percentage of CO-depleted BSSs is
small (of the order of $10\%$), thus indicating either that
CO-depletion is only temporarily visible on the BSS surface (and then
it is cleaned up by the subsequent evolution), or that the specific
formation channel generating this feature has a limited efficiency in
GCs.

A quite intriguing result seems instead to emerge from the measurement
of the BSS rotational velocities: in M4, we \cite{lovisi10} found that
$\sim 40\%$ of the measured BSSs are fast rotators, with rotational
velocities $v\sin i >50$\,km/s (see Fig.~ \ref{vrot_m4}). {\it This
  is the largest frequency of rapidly rotating BSSs ever detected in a
  GC.}  Interesting enough, while a similar fraction has been found in
$\omega$\,Cen (Mucciarelli et al. 2013, in prep.), significantly different
results have been found in 47\,Tuc, NGC\,6397 and M30, where almost
all ($92\%-94\%$) BSSs rotate slowly ($v\sin i <20$\,km/s \cite{fe06COdep, lovisi12}).  These results
suggest a possible correlation between the total fraction of rapidly
spinning BSSs and the density of the host cluster: in fact, GCs with
the largest fraction of fast rotators are also the loosest in our
sample ($\log\rho_0=3.91$ and 3.43 in units of M$_\odot$/pc$^3$
  for M4\index{M4} and $\omega$\,Cen\index{$\omega$ Centauri}, compared to $\log\rho_0=5.20$ for 47\,Tuc\index{47 Tucanae}
  and the very high central densities of the PCC clusters NGC\,6397\index{NGC 6397}
  and M30\index{M30}; \cite{mlvdm05}).  The total fraction of fast rotating BSSs in
the M4+$\omega$\,Cen sample is $\sim 33\%$, whereas it is only $\sim
4\%$ in the higher-density sample of 47\,Tuc+NGC\,6397+M30 (see
Fig. \ref{vrot_all}).  If confirmed, this would be the first evidence
of a direct link between the BSS physical properties and the
characteristics of the host cluster, and it could lead to interesting
scenarios never explored before.  These results indicate that some
braking mechanism (either magnetic braking or disk locking, as
proposed by Leonard \& Livio \cite {leoliv95} and Sills, Adams \& Davies \cite{sills05}, or something
different/additional) could somehow depend on the parent cluster
environment. For instance, recurrent stellar interactions might be
efficient in decreasing the BSS rotational velocities, while in loose
GCs, where stellar interactions are less frequent, the initial
rotational velocities of BSSs might be preserved and a larger fraction
of fast rotators should be observable.

While the mystery of BSS formation is not completely solved yet,
detailed photometric and spectroscopic observations of these puzzling
stars are providing crucial information about their physical
properties, also shedding new light on the global dynamical evolution
of stellar systems.

\begin{acknowledgement}
Most of the results discussed in this chapter have
been obtained within the project \emph{Cosmic-Lab} (PI: Ferraro, see {\tt
  http:://www.cosmic-lab.eu}), a 5-year project funded by the European
European Research Council under the 2010 \emph{Advanced Grant} call
(contract ERC-2010-AdG-267675).  We warmly thank the other team
members involved in this research: Giacomo Beccari, Paolo Miocchi, Mario
Pasquato, Nicoletta Sanna and Rodrigo Contreras Ramos. The authors
dedicate this chapter to the memory of Bob Rood, a pioneer in the theory
of the evolution of low mass stars and a dear friend who shared our
enthusiasm for the BSS topic and who unexpectedly passed away on 2
November 2011.
\end{acknowledgement}

\backmatter
\printindex



\begin{thebibliography}{}

\bibitem{anderson02} Anderson, J.: Omega
  Centauri, A Unique Window into Astrophysics, ASP 265, 87 (2002)

\bibitem{auriere90} Auriere, M., Lauzeral, C.,
  Ortolani, S.: Nature {\bf  344}, 638 (1990)

\bibitem{bai95} Bailyn, C. D.:  A\&ARA {\bf 33}, 133 (1995)

\bibitem{bec06} Beccari, G., Ferraro, F.~R.,
  Lanzoni, B.,  Bellazzini, M.: ApJL {\bf  652}, L121 (2006)

\bibitem{bec07} Beccari. G., et al.: ApJ {\bf 
  679}, 712 (2007)

\bibitem{beccari11pal14} Beccari, G., Sollima,
  A., Ferraro, F.~R., et al.: ApJL {\bf  737}, L3 (2011)

\bibitem{benz87} Benz, W.,  Hills, J. G.: 
  ApJ {\bf  323}, 614 (1987)

\bibitem{bolte93} Bolte, M., Hesser, J.~E., 
 Stetson, P.~B.: ApJL {\bf  408}, L89 (1993)

\bibitem{buon94} Buonanno, R., Corsi, C.E.,
  Buzzoni, A., Cacciari, C., Ferraro, F.R., Fusi Pecci, F.:
  A\&A {\bf  290}, 69 (1994)

\bibitem{cariulo03} Cariulo, P., Degl'Innocenti,
  S.  Castellani, V.: A\&A {\bf  412}, 1121 (2003)

\bibitem{carretta05} Carretta, E., Gratton,
  R. G., Lucatello, S., Bragaglia, A.,  Bonifacio, P.: A\&A {\bf 
  433}, 597 (2005)

\bibitem{contrerasm75} Contreras Ramos,
  R., Ferraro, F.~R., Dalessandro, E., Lanzoni, B.,  Rood,
  R.~T.: ApJ {\bf  748}, 91 (2012)

\bibitem{ema08_6388} Dalessandro, E., et
  al.: A\&A {\bf  677}, 1069 (2008a)

\bibitem{ema_2419} Dalessandro, E., et
  al.: A\&A {\bf  681}, 311 (2008b)
  
 \bibitem{dale2013} Dalessandro, E., 
Ferraro, F.~R., Massari, D., et al.: ApJ {\bf 778}, 135 (2013)


\bibitem{davies04} Davies, M.~B.,
  Piotto, G.,  de Angeli, F.: MNRAS {\bf 349}, 129 (2004)

\bibitem{demarco05} De Marco, O., et
  al.: ApJ {\bf  632}, 894 (2005)

\bibitem{djorgmey93} Djorgovski, S., 
  Meylan, G.: Structure and Dynamics of Globular Clusters, ASP 50,
  325 (1993)

\bibitem{djorgking86} Djorgovski, S., 
  King, I.~R.: ApJL {\bf  305}, L61 (1986)

\bibitem{dotter08} Dotter, A., Sarajedini, A., 
  Yang, S.-C.: AJ {\bf  136}, 1407 (2008)

\bibitem{feparesce93} Ferraro, F. R., 
  Paresce, F.: AJ {\bf  106},154 (1993)

\bibitem{fe93} Ferraro F. R., Pecci F. F.,
Cacciari C., Corsi C., Buonanno R., Fahlman G. G., Richer H. B.: 
AJ {\bf  106}, 2324 (1993)

\bibitem{fe95} Ferraro, F. R., Fusi Pecci, F.,
   Bellazzini, M.: A\&A {\bf  294}, 80 (1995)

\bibitem{fe97} Ferraro, F. R., et
al.: Paltrinieri, B., Fusi Pecci, F., Cacciari, C., Dorman, B., Rood,
R. T., Buonanno, R., Corsi, C. E., Burgarella, D.,  Laget, M.:
A\&A {\bf  324}, 915 (1997; F97)

\bibitem{fe99} Ferraro, F. R., Paltrinieri, B.,
  Rood, R. T.,  Dorman, B.: ApJ {\bf  522}, 983 (1999a)

\bibitem{fe03sixGCs} Ferraro, F. R., Sills,
A., Rood, R. T., Paltrinieri, B.,  Buonanno, R. :  ApJ {\bf  588},
464 (2003)

\bibitem{fe036752} Ferraro, F.~R., 
Possenti, A., Sabbi, E., et al.:  ApJ {\bf  595}, 179 (2003)

\bibitem{fe04} Ferraro, F. R., Beccari, G.,
Rood, R. T., Bellazzini, M., Sills, A.,  Sabbi, E.:  ApJ {\bf  603},
127 (2004)

\bibitem{fe06} Ferraro, F. R., Sollima, A.,
Rood, R. T., Origlia, L., Pancino, E.,  Bellazzini, M.:  ApJ {\bf 
638}, 433 (2006a)

\bibitem{fe06COdep} Ferraro, F. R., et
al.:  A\&A {\bf  647}, L53 (2006b)

\bibitem{fe09m30} Ferraro, F.~R., 
Beccari, G., Dalessandro, E., et al.: Nature {\bf 462}, 1028 (2009; F09)

\bibitem{fe12} Ferraro, F.~R., 
Lanzoni, B., Dalessandro, E., et al.: Nature {\bf 492}, 393 (2012; F12)

\bibitem{fp92} Fusi Pecci, F., Ferraro,
F. R., Corsi, C. E., Cacciari, C.,  Buonanno, R.: ApJ {\bf  104},
1831 (1992)

\bibitem{guha94} Guhathakurta, P., Yanny,
  B., Bahcall, J. N., Schneider, D. P.: ApJ {\bf  108}, 1786
(1994)

\bibitem{gilliland98} Gilliland, R.~L., Bono,
G., Edmonds, P.~D., et al.: ApJ {\bf  507}, 818 (1998)

\bibitem{heggie08m4} Heggie, D.~C., 
  Giersz, M.: MNRAS {\bf 389}, 1858 (2008)

\bibitem{hillsday76} Hills, J. G.,  Day,
  C. A.: ApJL {\bf 17}, 87 (1976)

\bibitem{jordi09} Jordi, K., Grebel, E.~K.,
  Hilker, M., et al.: AJ {\bf  137}, 4586 (2009)

\bibitem{knigge09} Knigge, C., Leigh, N., 
  Sills, A.: Nature {\bf 457}, 288 (2009)

\bibitem{lan07M5} Lanzoni, B., Dalessandro,
E., Ferraro, F. R., Mancini, C., Beccari, G., Rood, R. T., Mapelli,
M.,  Sigurdsson, S. : ApJ {\bf  663}, 267 (2007a)

\bibitem{lan071904} Lanzoni, B., et
  al.: A\&A {\bf  663}, 1040 (2007b)

\bibitem{lan07M55} Lanzoni, B., et al.:
  ApJ {\bf  670}, 1065 (2007c)

\bibitem{lan07imbh} Lanzoni, B., 
Dalessandro, E., Ferraro, F.~R., et al.: ApJL {\bf  668}, L139 (2007)

\bibitem{lan13} Lanzoni, B., Mucciarelli, A.,
  Origlia, L., et al.: ApJ {\bf  769}, 107 (2013)

\bibitem{leigh07} Leigh, N., Sills, A., 
 Knigge, C.: ApJ {\bf  661}, 210 (2007)

\bibitem{leigh11} Leigh, N., Sills, A., 
 Knigge, C.: MNRAS {\bf 415}, 3771 (2011)

\bibitem{leigh13} Leigh, N., Knigge, C., 
Sills, A., et al.: MNRAS {\bf 428}, 897 (2013)

\bibitem{leoliv95} Leonard, P. J. T.,  Livio,
  M.:  ApJL {\bf  447}, 121 (1995)

\bibitem{lomb95} Lombardi, J. C. Jr., Rasio,
  F. A.,  Shapiro, S. L.: ApJL {\bf  445}, 117 (1995)

\bibitem{lovisi10} Lovisi, L., Mucciarelli, 
A., Ferraro, F.~R., et al.: ApJL {\bf  719}, 121 (2010)

\bibitem{lovisi12} Lovisi, L., Mucciarelli, A.,
  Lanzoni, B., et al.: ApJ {\bf  754}, 91 (2012)

\bibitem{luc03}Lucatello, S., Gratton,
  R.G.: A\&A {\bf  406}, L691 (2003)

\bibitem{mapelli04} Mapelli, M., Sigurdsson,
S., Colpi, M., Ferraro, F. R., Possenti, A., Rood, R. T., Sills,
A.,  Beccari, G.:  ApJL {\bf  605}, 29 (2004)

\bibitem{mapelli06} Mapelli, M., Sigurdsson,
S., Ferraro, F. R., Colpi, M., Possenti, A.,  Lanzoni, B.:
MNRAS {\bf 373}, 361 (2006)

\bibitem{mapelli09} Mapelli, M., Ripamonti, E.,
  Battaglia, G., et al.: MNRAS {\bf 396}, 1771 (2009)

\bibitem{marinfranch09}
  Mar{\'{\i}}n-Franch, A., Aparicio, A., Piotto, G., et al.:
  ApJ {\bf  694}, 1498 (2009)

\bibitem{mathieu09} Mathieu, R.~D., 
  Geller, A.~M.: Nature {\bf 462}, 1032 (2009)

\bibitem{mcrea64} McCrea, W. H.: MNRAS {\bf 128}, 147 (1964)

\bibitem{mlvdm05} McLaughlin,
  D.~E.,  van der Marel, R.~P.: A\&A {\bf 161}, 304 (2005)

\bibitem{meylanheggie97} Meylan, G., 
  Heggie, D.~C.: A\&A Rev. {\bf 8}, 1 (1997)

\bibitem{milone12} Milone, A.~P., Piotto, G.,
  Bedin, L.~R., et al.: A\&A {\bf  540}, A16 (2012)

\bibitem{monelli12} Monelli, M., Cassisi, S.,
  Mapelli, M., et al.: ApJ {\bf  744}, 157 (2012)

\bibitem{moretti08} Moretti, A., de Angeli, F.,
   Piotto, G.: A\&A {\bf  483}, 183 (2008)

\bibitem{paresce91}Paresce, F., Meylan, G.,
  Shara, M., et al.:  Nature {\bf  352}, 297 (1991)

\bibitem{pietrinferni06} Pietrinferni, A.,
  Cassisi, S., Salaris, M.,  Castelli, F.: ApJ {\bf  642}, 797 (2006)

\bibitem{piotto04} Piotto, G., et al.: 
  ApJ {\bf  604}, L109 (2004)

\bibitem{quievy09} Quievy, D., Charbonneau, P.,
  Michaud, G.,  Richer, J.: A\&A {\bf  500}, 1163 (2009)

\bibitem{renzfusi88} Renzini, A.,  Fusi
  Pecci, F.: A\&ARA {\bf 26}, 199 (1988)

\bibitem{sabbi04} Sabbi, E., Ferraro, F. R., Sills, A.,
 Rood, R. T.: ApJ {\bf  617}, 1296 (2004)

\bibitem{sand53} Sandage A. R.: AJ {\bf  58}, 61 (1953)

\bibitem{saraj93} Sarajedini, A.: Blue
  Stragglers, ASP {\bf 53}, 14 (1993)

\bibitem{saraj07} Sarajedini, A., Bedin,
  L.~R., Chaboyer, B., et al.: AJ {\bf  133}, 1658 (2007)

\bibitem{sarna96} Sarna, M. J.,  de Greve,
  J. P.: QJRAS {\bf 37}, 11 (1996)

\bibitem{shara97} Shara, M. M., Saffer, R. A., 
  Livio, M.: ApJ {\bf  489}, L59 (1997)

\bibitem{sigurd94}Sigurdsson, S., Davies,
  M.B.,  Bolte, M.: ApJL {\bf  431}, 115 (1994)

\bibitem{sills05} Sills, A., Adams, T.,  Davies,
  M.~B.: MNRAS {\bf 358}, 716 (2005)

\bibitem{sills09} Sills, A., Karakas, A., 
  Lattanzio, J.: ApJ {\bf  692}, 1411 (2009)

\bibitem{sollima08} Sollima, A. et al.:
  A\&A {\bf  481}, 701 (2008)

\bibitem{tian06} Tian, B., Deng, L., Han, Z., 
  Zhang, X.~B.: A\&A {\bf  455}, 247 (2006)

\bibitem{w06} Warren, S. R., Sandquist, E. L.,
   Bolte, M.: ApJ {\bf  648}, 1026 (2006)

\bibitem{zaggia97} Zaggia, S. R., Piotto, G. 
  Capaccioli M.: A\&A {\bf  327}, 1004 (1997)

\bibitem{zinnsearle76} Zinn, R.,  Searle, L.: 
  ApJ {\bf  209}, 734 (1976)

\end{thebibliography}
\end{document}